\documentclass[reprint,nofootinbib,superscriptaddress]{revtex4-2}
\usepackage[dvipsnames]{xcolor}
\usepackage{graphicx}
\usepackage{changepage} % Required for the adjustwidth environment
\usepackage{caption} % Required for the \captionof command
\usepackage{multirow} % Required for the \multirow command
\usepackage{hyperref}
\usepackage{amsmath}
\usepackage{amsfonts}
\usepackage{amssymb}
\usepackage{float}
\usepackage{caption}
\usepackage{soul}
\usepackage{upgreek}
\usepackage[normalem]{ulem}

\usepackage{ragged2e}

\DeclareCaptionJustification{justified}{\justifying}

\usepackage[capitalise,noabbrev,nameinlink]{cleveref}
\hypersetup{
citecolor=blue,
 colorlinks=true,
 linkcolor=blue,
 filecolor=magenta,
 urlcolor=cyan,
}

\graphicspath{{./figures/}}

\Crefname{equation}{Eq.}{Eqs.}

\begin{document}

\title{Modern nuclear and astrophysical constraints of dense matter in a redefined chiral approach}

\author{Rajesh Kumar}
\email{rkumar6@kent.edu}
\affiliation{Center for Nuclear Research, Department of Physics, Kent State University, Kent, OH 44243 USA}

\author{Yuhan Wang}
% \email{ywang156@kent.edu}
\affiliation{Center for Nuclear Research, Department of Physics, Kent State University, Kent, OH 44243 USA}

\author{Nikolas Cruz Camacho}
\affiliation{Illinois Center for Advanced Studies of the Universe, Department of Physics, University of Illinois at Urbana-Champaign, Urbana, IL 61801, USA}

\author{Arvind Kumar}
% \email{kumara@nitj.ac.in}
\affiliation{Department of Physics, Dr. B R Ambedkar National Institute of Technology Jalandhar, Punjab, 144008  INDIA}

\author{Jacquelyn Noronha-Hostler}
\affiliation{Illinois Center for Advanced Studies of the Universe, Department of Physics, University of Illinois at Urbana-Champaign, Urbana, IL 61801, USA}

\author{Veronica Dexheimer}
\email{vdexheim@kent.edu}
\affiliation{Center for Nuclear Research, Department of Physics, Kent State University, Kent, OH 44243 USA}

\date{\today}

\begin{abstract}
We explore the Quantum Chromodynamics (QCD) phase diagram's complexities, including quark deconfinement transitions, liquid-gas phase changes, and critical points by using the chiral mean-field (CMF) model that is  able to capture all these features. We introduce a vector meson field redefinition within the CMF framework, enabling precise adjustments of meson masses and coupling strengths related to vector meson interactions. Performing a new fit to the  deconfinement potential, we are able to replicate  recent lattice QCD results, low energy nuclear physics properties, neutron star observational data, and key phase diagram features as per modern constraints. This approach enhances our understanding of vector mesons' roles in mediating nuclear interactions and their impact on the equation of state, contributing to a more comprehensive understanding of the QCD phase diagram and its implications for nuclear and astrophysical phenomena.
\end{abstract}

\maketitle

%\newpage
%\tableofcontents
%\newpage

% \keywords{neutron star \and equation of state \and quark matter}

\maketitle

\section{Introduction}

% High energy matter
    Hot and/or dense Quantum Chromodynamics (QCD) matter is a fascinating area of research and its understanding requires knowledge of  theoretical and experimental nuclear physics, astrophysics,  particle physics, and gravity \cite{Dexheimer:2020zzs,Lovato:2022vgq,Sorensen:2023zkk}. It includes the extreme conditions of temperature  that existed shortly after the Big Bang, during the early moments of the universe's formation. These conditions  are believed to be reproduced in relativistic particle collisions, such as those created in high energy particle accelerators like the Large Hadron Collider (LHC) and the Relativistic Heavy-Ion Collider (RHIC)~\cite{Heinz:2000ba,Vogt:2007zz}. On the other hand, QCD matter at effectively zero temperature (in the range of MeV) in neutron stars is another fascinating and complex area of study within nuclear astrophysics  \cite{Baym:2017whm, Alford:2007xm}. Neutron stars are incredibly dense celestial objects  formed when massive stars undergo  supernova explosions at the end of their life cycles. Significant interest is focused on trying to find exotic degrees of freedom like hyperons or deconfined quarks in the core of neutron stars \cite{Lattimer:2021emm,Burgio:2021vgk,Annala:2019puf}, since this would be the only regime in the universe where they could be stable. 
    
% Phase diagrams

\begin{figure}[t!]
\includegraphics[scale=0.33]{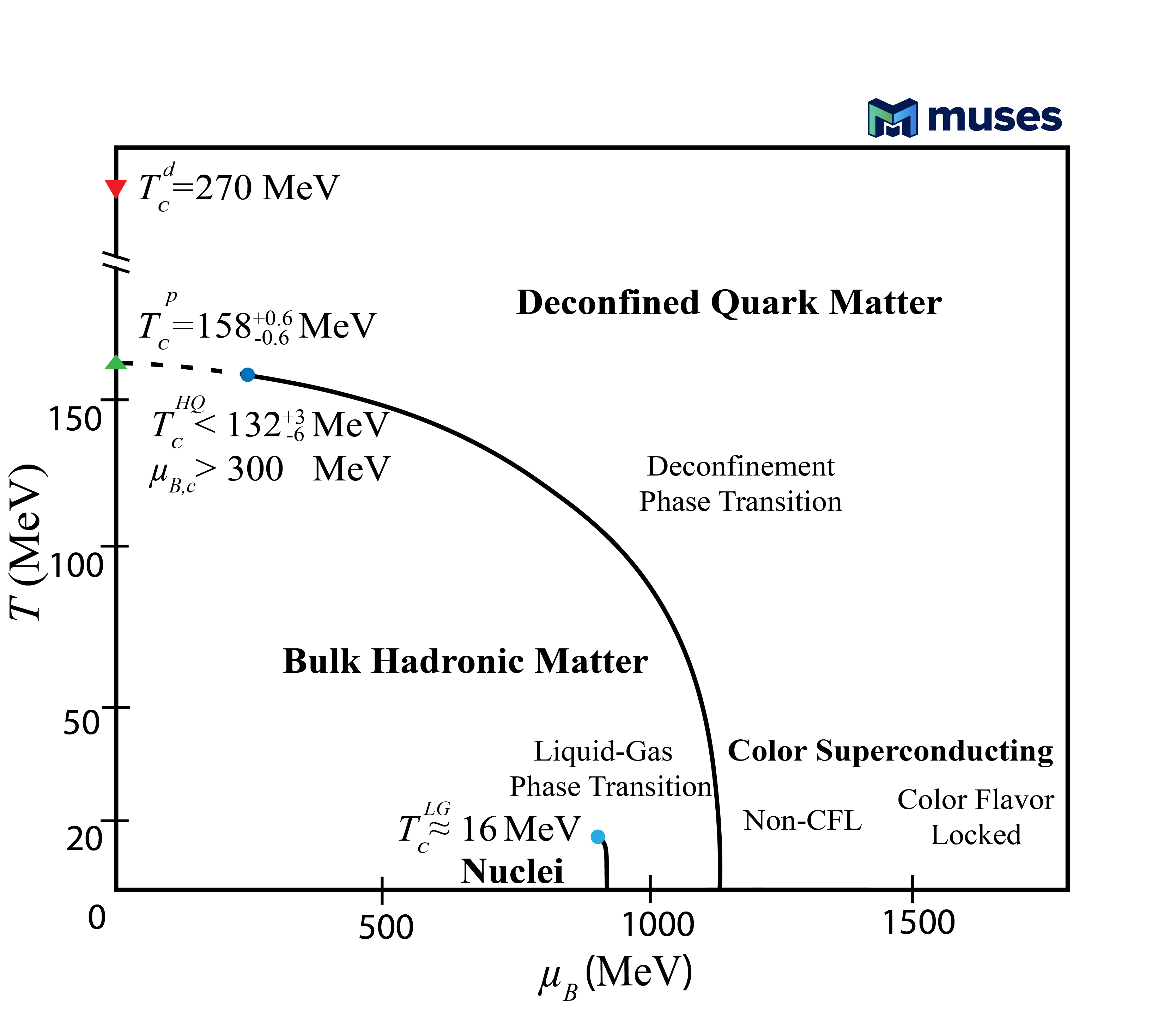}
\caption{A rough sketch of the QCD phase diagram  showing different phases, lattice QCD results, and experimental data. $T^d_c$ and $T^p_c$ denote the deconfinement phase transition and pseudo critical phase transition respectively (both at $\mu_B=0$) while $\mu_{B,c}$ and $T^{HQ}_c$ represent the critical baryon chemical potential and critical temperature for hadron quark phase transition, and $T^{LG}_c$ the critical temperature for liquid-gas phase transition.} 
\label{fig:QCDPD}
\end{figure}

The QCD phase diagram delineates phases of strongly interacting matter, usually under varying temperature ($T$) and baryon chemical potential ($\mu_B$). At low $T$ and $\mu_B$, quarks and gluons are confined within hadrons (hadronic phase) and are expected to  transition to an effective liberated state called deconfined quark matter at high $T$ and/or $\mu_B$~ (see recent reviews \cite{Philipsen:2012nu,Ratti:2018ksb} from lattice QCD).   
In addition to the confinement/deconfinement quark hadron phase transition,   there also exists a phase transition from nuclei to bulk hadronic matter known as liquid-gas phase transition at   \mbox{$T^{\rm LG}_c\simeq 15-17$ MeV}\mbox{($ \mu^{\rm LG}_{B,c}\approx910$ MeV)}~\cite{Natowitz:2002nw,Karnaukhov:2008be,Elliott:2013pna}. The QCD phase diagram is believed to encompass two critical points,  the liquid-gas and hadron-quark.  In both cases, the first-order phase transition coexistence lines are thought to end at the respective critical points and become indistinct after that, which is referred to as a crossover regime (see \Cref{fig:QCDPD}).
Lattice QCD has proven to be highly effective for investigating strong interactions in the vicinity of and beyond the deconfinement phase transition zone within the QCD phase diagram in the high $T$ and low $\mu_B$ regime, primarily due to its ability to handle non-perturbative aspects~\cite{Ratti:2018ksb}. 
Based on the latest lattice results, no sign of critical behavior has been found up to $\mu_B\approx300$ MeV~\cite{Borsanyi:2020fev,Borsanyi:2021sxv}, and the  critical temperature  is expected to be smaller than $T^{\rm HQ}_c=132^{+3}_{-6}$ MeV for isospin symmetric matter with zero baryon ($\mu_B$), charge ($\mu_Q$) and strange ($\mu_S$) chemical potential~\cite{HotQCD:2019xnw}. Within  lattice QCD results, the crossover or pseudo-critical temperature (at $\mu_B$=0 axis) has been identified with extreme accuracy as $T^p_c=158\pm0.6$ MeV~\cite{Borsanyi:2020fev}, in addition to a first order deconfinement phase transition  for pure  glue (without quarks)  at a temperature of $T_c^d=$ 270 MeV~\cite{Roessner:2006xn}. On the other side of the diagram, in neutron stars, the critical density $n^d_{B,c}$, which marks the initial stage of the transition from hadronic matter to quark deconfinement  is still not yet well constrained.

%Low energy nuclear physics

A core requirement for dense matter theories is the accurate reproduction of experimental data for isospin-symmetric nuclear matter at low temperature and around nuclear saturation density $n_{\mathrm{sat}}$. This entails crucial observables, such as the binding energy per nucleon $B/A$,  compressibility $K$,  symmetry energy $E_{\rm sym}$, and  slope parameter $L$. Notably, recent progress has been made in the measurement of the parity-violating asymmetry term $A_{\mathrm{PV}}$ through elastic scattering of longitudinally polarized electrons on $^{208}$Pb. With this, the PREX collaboration's findings have facilitated the determination of the nuclear saturation density value $n_{\mathrm{sat}}=0.1480\pm0.0038~\mathrm{fm}^{-3}$ \cite{PREX:2021umo}. The binding energy per nucleon values were determined to be $B/A=$-15.677 MeV at a saturation density of $n_{\mathrm{sat}}$ = 0.16146 fm$^{-3}$~\cite{Myers:1966zz} and $B/A=$-16.24 MeV at $n_{\mathrm{sat}}$ = 0.16114 fm$^{-3}$~\cite{Myers:1995wx}. These values were obtained by analyzing experimental data from heavy nuclei masses and ground state masses of nuclei with neutron ($N$) and proton ($Z$) numbers greater than or equal to 8, respectively.  The Isovector Giant Monopole Resonance (ISGMR) collective nucleon excitations from nuclei such as $^{90}$Zr and $^{208}$Pb have suggested a value of $K=240\pm20$ MeV for the incompressibility of infinite nuclear matter \cite{Colo:2013yta,Todd-Rutel:2005yzo,Colo:2004mj,Agrawal:2003xb}. But note that, in a comprehensive review \cite{Stone:2014wza}, various methodologies and theories used between 1961 and 2016 led to a much larger range of $K$ values, from 100 MeV to 380 MeV, with relativistic mean-field models often predicting higher values. Finally, a range of \mbox{$250 < K< 315$ MeV} was obtained without assuming any specific microscopic model, except for the Coulomb effect~\cite{Stone:2014wza}.

% Asymmetric matter

Going further, the symmetry energy $E_{\rm sym}$ is the energy (per baryon) difference between nuclear matter with equal numbers of protons and neutrons (isospin-symmetric) and pure neutron matter. The slope parameter ($L$)  is a measure of how rapidly  $E_{\rm sym}$ (at $n_{\mathrm{sat}}$) changes with the baryon density. Both $E_{\rm sym}$ and $L$ are important quantities for understanding various nuclear phenomena, such as neutron star properties and low-energy heavy-ion collisions~\cite{Yao:2023yda}. In Ref.\ \cite{Li:2019xxz}, a comprehensive assessment based on 28 model evaluations utilized terrestrial nuclear experiments and astrophysical data to determine $E_{\rm sym}$ and $L$ at saturation density. Fiducial values emerged as $(31.6\pm2.7)$ MeV for $E_{\rm sym}$ and $(58.9\pm16)$ MeV for $L$. Extracting $E_{\rm sym}$ from experimental nuclear masses yielded $L=(50.0\pm15.5)$ MeV at $n_{\mathrm{sat}}=0.16$ fm$^{-3}$ \cite{Fan:2014rha}. Interestingly, addressing $^{208}$Pb's neutron skin thickness, PREX-II constrained the symmetry energy, revealing a large slope $L=(106\pm37)$ MeV~\cite{Reed:2021nqk}, consistently exceeding current bounds. On the other hand, another PREX-II study examined the parity-violating asymmetry $A_{\text{PV}}$ for $^{208}$Pb, leading to a neutron skin thickness $R^{208}_{\text{skin}}=(0.19\pm0.02)$ fm and a much smaller value of slope $L=(54\pm8)$ MeV~\cite{Reinhard:2021utv}, consistent with prior astrophysical estimates. Furthermore,  the hyperon potential ($U_H$)  describes the interactions between hyperons and nucleons  at $n_{\rm{sat}}$ for isospin-symmetric matter. The $\Lambda$-nucleon potential, obtained from 1980s experiments, is firmly negative at around $U_\Lambda \sim -28$ MeV~\cite{Millener:1988hp}, with recent estimates clustering between $-32$ and $-30$ MeV~\cite{Fortin:2017dsj}. The measurements from KEK Japan  indicate a repulsive potential for $\Sigma$ ($U_\Sigma = 30 \pm 20$ MeV) and a joint collaboration between  KEK and JPARC Japan  give an attractive potential for $\Xi$ ($U_\Xi = -21.9 \pm 0.7$ MeV)~\cite{Gal:2016boi}. The ALICE collaboration's p-$\Xi$ correlation functions report a less attractive potential of $U_\Xi = -4$ MeV, aligning with HAL-QCD collaboration's (2+1)D lattice QCD calculations, yielding $U_\Xi = -4$ MeV, $U_\Lambda = -28$ MeV, and $U_\Sigma = +15$ MeV, with a statistical error of approximately $\pm 2$ MeV~\cite{Fabbietti:2020bfg,ALICE:2020mfd,Inoue:2019jme}.

The nuclear matter characteristics exhibit significant correlations with macroscopic observables of neutron stars, such as the maximum mass ($M_{\rm max}$), radius $R_{M_{\rm max}}$, and tidal deformability ($\tilde{\Lambda}$). The determination of neutron star radii from NICER's X-ray observations yield values of $12.39_{-1.98}^{+1.30}$ km for a $2.072_{-0.066}^{+0.067}$ solar mass ($M_\odot$)\cite{Miller:2019cac}  and $13.7_{-1.5}^{+2.6}$ km for a $2.08_{-0.07}^{+0.07}$ $M_\odot$\cite{Riley:2019yda} neutron star, respectively. Additionally, the gravitational wave event GW170817, resulting from the merger of binary neutron stars (BNS), imposes a constraint on the tidal deformability, indicating $\tilde{\Lambda} < 800$ for neutron stars with a mass of 1.4 solar masses~\cite{LIGOScientific:2018cki}. A more detailed discussion about constraints from first principles, low energy nuclear experiments, heavy-ion experiments, and astrophysical observations is given in our recent review article~\cite{MUSES:2023hyz}.

% Effective models
 
Solving QCD analytically is a complex task, despite a well-defined Lagrangian. Lattice QCD  represents space-time on a lattice where quarks reside at the vertices, connected by gluon lines \cite{Troyer:2004ge}. It works very well at $\mu_B=0$ axis but cannot be applied directly to finite $\mu_B$ region due to the sign problem~\cite{Muroya:2003qs,deForcrand:2010ys}. However, the Taylor and alternative expansion schemes at $\mu_B=0$, enable the derivation of the lattice QCD Equation of State (EoS) up to a chemical potential $\mu_B\sim 3.5\,T$ using expansion coefficients at $\mu_B=0$~\cite{Borsanyi:2021sxv,Borsanyi:2022qlh}.  The Polyakov loop is a gauge-invariant quantity that can be used to characterize the behavior of quarks in the presence of a thermal bath. At $\mu_B=0$, the scalar field $\Phi$ associated with the Polyakov loop serves as an order parameter for $Z(3)$ symmetry in the context of pure gluonic interactions. With quarks included, the transition from the confined phase to the deconfined phase becomes a crossover rather than a sharp phase transition. This means that the behavior of the Polyakov loop may show gradual changes rather than a sudden jump, indicating a smoother transition from confined hadronic matter to the deconfined QGP phase. Therefore, it acts as an approximate order parameter when quarks are added~\cite{Roessner:2006xn}. Perturbative QCD (pQCD) is applicable at large $\mu_B$ and/or $T$, but breaks down near the deconfinement phase transition due to large coupling constants \cite{Andersen:2002jz,Fraga:2013qra,Kurkela:2016was}. At high temperature and low $\mu_B$, resummed pQCD calculations are in agreement with lattice data for  \mbox{$T\geq250$ MeV}~\cite{Andersen:1999fw,Andersen:1999sf,Andersen:1999va,Haque:2014rua,Haque:2020eyj}. Chiral effective theory ($\chi$EFT) is suitable at low densities and temperatures~\cite{Tews:2012fj}. Despite these methods, the QCD phase diagram remains largely uncharted (see Figure 1 of Ref.~\cite{MUSES:2023hyz}). That is where effective models come in, to bridge the gap between QCD complexities and first-principle limitations, providing valuable insights across a broad spectrum of QCD phenomena and constructing Lagrangians with the appropriate degrees of freedom~\cite{Walecka1974,Nambu:1961tp,Fukushima:2003fw}. In particular, relativistic chiral mean-field models can reproduce the restoration of chiral symmetry and quantify how hadronic masses are influenced by the medium \cite{Schaffner-Bielich:1998mra,Weinberg:1968de,Walecka1974,Gallas:2009qp,Kovacs:2021ger}. 

% CMF

From the latter class,  non-linear chiral  models  stand out, based on a non-linear realization of chiral symmetry \cite{Weinberg:1968de,Papazoglou:1998vr,Bonanno:2008tt,Kumar:2018ujk,Motornenko:2019arp}.  The introduction of  a Polyakov loop-inspired potential  in a  non-linear chiral model as a mechanism to deconfine quarks gave rise to the chiral mean-field (CMF) model~\cite{Dexheimer:2009hi}. Within the mean-field approximation,  the  CMF model agrees well with nuclear data~\cite{Dexheimer:2008ax}. It offers a unified description, allowing one to  investigate the properties of strongly interacting matter  in heavy-ion collisions and compact stars, integrating quark deconfinement through an order parameter $\Phi$ with values dependent upon a Polyakov loop-like potential~\cite{Ratti:2006ka,Roessner:2006xn}. The CMF model accommodates various temperatures, densities, and magnetic fields \cite{Dexheimer:2011pz,Franzon:2015sya,Dexheimer:2021sxs,Marquez:2022fzh,Peterson:2023bmr}, enabling it to be used to explore various regions of QCD phase diagram \cite{Dexheimer:2009hi,Hempel:2013tfa,Roark:2018uls,Aryal:2020ocm,Peterson:2023bmr}. 
However, these past works did not include a consistent treatment of mesons.  The mesons lacked in-medium contributions and the vector-mesons masses were degenerate.

% Vector mesons

Understanding the importance of vector meson masses and interactions is a necessary step in the direction of incorporating thermal mesons in the formalism. In chiral models, hadronic masses are generated by interactions with the medium and can depend on $T$, $\mu_B$, etc. In  relativistic mean-field models, vector interactions play a significant role in describing the behavior of hadrons and their connection within the framework of QCD. For example,  vector mesons (such as the $\omega$ meson) couple to nucleons and interact with other hadrons, which further   play a significant role in determining the stiffness of the  EoS of nuclear matter in heavy-ion collisions and neutron stars. The role of vector mesons has been extensively studied in different theoretical approaches to determine properties of nuclear matter and compact stars~\cite{Lenzi:2023nox,Ferreira:2020kvu,Lourenco:2021lpn,Pisarski:2021aoz,Lopes:2021zfe,Ma:2022fmu,Li:2022okx,Sun:2022yor,Pradhan:2022txg,Thakur:2022dxb,Kubis:2023qfz,CamaraPereira:2016chj,Baym:2017whm,Wu:2018kww,Malfatti:2019tpg,Lopes:2020rqn,Alaverdyan:2020xnv,Cao:2020zxi,Otto:2020hoz,Marczenko:2020jma,Benic:2014jia,Miyatsu:2022wuy,Singh:2014zza, Lopes:2019shs, Kumari:2020mci,Kumar:2020vys,Dexheimer:2015qha,Dexheimer:2018dhb,Dexheimer:2020rlp,Ye:2023fhy,Cierniak:2018aet,Kojo:2019raj,Shahrbaf:2019vtf,Wu:2020knd,Sun:2020bbn,Huang:2022jiu,Kumar:2023lhv,Miyatsu:2022wuy,Li:2022okx,Aguirre:2022zxn}.

% Vector mesons in CMF

In the hadronic non-linear chiral model ~\cite{Papazoglou:1998vr,Zschiesche:2003qq,Kumar:2018ujk,Dexheimer:2008ax}, vector mesons ($\omega$, $\rho$ and $\phi$) were also introduced as mediators of the strong interaction between nucleons and hyperons. 
The degenerate masses of different vector mesons ($\omega$, $\rho$, and $\phi$) were broken by introducing a renormalization of vector mesons through the utilization of proper invariants \cite{Papazoglou:1998vr,Zschiesche:2003qq}.  For finite nuclei,  the renormalization  of vector mesons was used to break the mass degeneracy of  $\omega$, $\phi$ and $K^*$ mesons   \cite{Papazoglou:1998vr}. In particular,  in Ref. \cite{Zschiesche:2003qq}, utilizing a combination of two invariants, vector meson renormalization  was employed to lift the mass degeneracy among $\omega$, $\phi$, and $\rho$ mesons. Note that, the renormalization  of vector meson in chiral models involves adjusting parameters related to the coupling strengths or masses of the vector mesons to achieve a better match between the model predictions and experimental data. This is a complex and iterative process, often requiring sophisticated computational techniques and comparisons with experimental observables. 

% This work

 In the present study, the term ``renormalization" is replaced by  ``field redefinition" due to its potential confusion with the renormalization method aimed at addressing divergences in pQCD calculations. We  employ vector meson  field redefinition  to break the mass degeneracy between the vector mesons in the CMF model for the first time. We refit the vector meson coupling strengths to nucleons such as $g_{N\omega }$, $g_{N\rho }$ and $g_4$ (the coupling strength related to the effective self-interactive vector Lagrangian) to the saturation properties of the nuclear matter. The addition of a vector meson  field redefinition then significantly affects other properties within the CMF model, such that we need to reparametrize other parameters that we detail here. These changes then require that we must refit the coupling constants related to the Polyakov loop-like  potential within the CMF model  to reproduce recent lattice data. We also incorporate  updated information about the phase diagram that has changed since the last time the finite temperature CMF model parameters were parameterized (in 2008). The changes include state-of-the-art and updated information about the  deconfinement phase transition,  pseudo-critical temperature, liquid-gas critical point,  deconfinement critical point, and observational data for neutron stars. Note that the field redefined  vector mesons significantly affect the in-medium properties of vector mesons, which will be studied in a future work.

% Outline of the paper

The outline of this paper is as follows. In \cref{sec:CMF}, the details of the CMF model are given along with the Polyakov loop-like potential.  In \cref{sec:Lag_vec_ren}, a detailed derivation of  field redefined  vector meson  is provided and the same is applied for different self-interactions of vector mesons. In \cref{sec:RnD}, the  results are presented for each part of QCD phase diagram. Finally, we present a summary  with discussions in \cref{sec:conclusion}.

\section{Formalism}
\label{sec:formalism}
\subsection{Chiral mean-field model}
\label{sec:CMF}

In this work, we build on the CMF model, which incorporates fundamental QCD aspects like the trace anomaly, spontaneous breaking of chiral symmetry  and deconfinement \cite{Weinberg:1968de,Coleman:1969sm}. Based on a non-linear realization of chiral symmetry, this framework employs scalar and vector fields to describe meson-baryon/quark interactions. The  scalar-isoscalar field $\sigma$ corresponds loosely to the light quark composed meson $\sigma_0(500)(u\bar{d})$. A strange scalar-isoscalar field $\zeta$ is linked to the strange quark-containing meson $s\bar{s}$, crucial to describe strange matter \cite{Zakout:1999qu}. Additionally, the scalar-isovector field $\delta$ addresses isospin asymmetric matter and introduces mass splitting between isospin multiplet and being  associated with the meson $(\bar u u-\bar d d)$ \cite{Kubis:1997ew,Hofmann:2000vz}. These fields mediate interactions among nucleons, hyperons, and quarks, contributing to attractive medium-range forces (scalar fields) and short-range repulsion (vector fields, e.g., vector-isoscalar $\omega$, strange  vector-isoscalar $\phi$, and vector-isovector  $\rho$) depending on $T,\mu_B$, etc. The scalar dilaton field, $\chi$, representing the hypothetical glueball field, is introduced to replicate QCD's trace anomaly \cite{Papazoglou:1998vr}. Nevertheless, due to the little overall contribution of $\chi$ field to  baryon thermodynamic quantities, we use the so-called frozen glueball approximation ($\chi=\chi_0$), where $\chi_0$ is the vacuum value of the dilaton field. 

The mean field approximation (MFA)   involves replacing the meson fields with their respective expectation values, effectively treating them as classical fields. As a result, only mesons along the diagonal of the scalar meson matrix $X$ (\Cref{Xmatrix_mfa})  have non-zero values due to the preservation of parity. Furthermore, all scalar and vector mesons are simplified into constants that are independent of both time and space. As a result of this approximation, the mean-field  CMF Lagrangian  reads \cite{Dexheimer:2009hi}
\begin{equation}
\mathcal{L}_{\rm CMF}=\mathcal{L}_{\rm
kin}+\mathcal{L}_{ \rm int}+\mathcal{L}_{\rm 
 scal}+\mathcal{L}_{\rm  vec}+\mathcal{L}_{\rm  esb} - U_{  \Phi}\,.
\label{eq:LCMF}
\end{equation}
Above, ${\cal L}_{\rm kin}$ stands for the  kinetic energy of spin-1/2 fermions (octet baryons + quarks), ${\cal L}_{\rm  int}$ represents  interactions of spin-1/2 fermions with vector and scalar mesons,  ${\cal L}_{\rm  scal}$ stands for the self-interactions of scalar mesons, while  ${\cal L}_{\rm  vec}$ contributes to vector meson masses and includes quartic self-interaction terms (see \Cref{sec:Lag_vec_ren} for details).   ${\cal L}_{\rm esb}$  denotes an explicit chiral symmetry breaking contribution  with the second term  (${\cal L}_{esb}$ of \cref{eq:detailed_Lag}) allowing the CMF model  to reproduce the experimental values of hyperon potentials and $U_{\Phi}$ denotes the deconfinement potential. Explicitly, these terms can be written as 
\begin{align}
\mathcal{L}_{\rm  kin}&=\sum_{i \, \in \, \mathrm{fermions}}\bigg[\bar{\psi}_ii\gamma_\mu\partial^\mu\psi_i\bigg], \nonumber \\
\mathcal{L}_{\rm  int}&=-\sum_{i  \in  \mathrm{fermions}}
\bar{\psi}_i\big[\gamma_0 \big(g_{i\omega}\omega+g_{i\rho}\rho+g_{i\phi}\phi\big) +m_i^*\big]\psi_i\,, \nonumber \\
\mathcal{L}_{\rm  scal}&=-\frac{1}{2}k_0\chi_0^2(\sigma^2+\zeta^2+\delta^2)+k_1(\sigma^2+\zeta^2+\delta^2)^2\nonumber \\
&+k_2\left[\frac{\sigma^4+\delta^4}{2}+\zeta^4+3 \left(\sigma\delta \right)^2\right]+k_3\chi_0\left(\sigma^2-\delta^2\right)\zeta\nonumber\\ &-k_4\chi_0^4\nonumber+\frac{\epsilon}{3}\chi_0^4\ln\left[\frac{\left(\sigma^2-\delta^2 \right)\zeta}{\sigma_0^2\zeta_0}\right], \nonumber \\
\mathcal{L}_{\rm  vec}&= \quad \mathrm{discussed}\ \mathrm{in}\ \mathrm{\cref{sec:Lag_vec_ren}},
 \nonumber \\
\mathcal{L}_{\rm  esb}&=-\left[m_{\pi}^{2}f_{\pi}\sigma+\left(\sqrt{2}m_{K}^{2}f_{K}-\frac{1}{\sqrt{2}}m_{\pi}^{2}f_{\pi}\right)\zeta\right] \nonumber \\
 &-m_3\sum_{i \, \in \, \mathrm{hyperons}}\bigg[\bar{\psi}_i  \left( \sqrt{2} (\sigma-\sigma_0) +  (\zeta-\zeta_0) \right)  \psi_i\bigg],
 \label{eq:detailed_Lag}
 \end{align}
 and
 
 \begin{align}
U_{\Phi}&=\left(a_0T^4+a_1\mu_B^4+a_2T^2\mu_B^2\right)\Phi^2\nonumber \\ &+a_3T_0^4\ln\left(1-6\Phi^2+8\Phi^3-3\Phi^4\right),
\label{eq:U_Phi}
\end{align}

where $\psi$ represents the fermionic field,  $g's$ denote the corresponding  coupling constants of fermions with  meson mean-fields, $k's$ are the fitting parameters associated with the scalar mesons, and $\epsilon$ is a model parameter related to the QCD trace anomaly. The variables: $m_K$, $m_\pi$, $f_K$, and $f_\pi$ are the masses and decay constants of the kaons and pions, respectively. The parameter $m_3$ is associated with the explicit chiral symmetry breaking and is fitted to reproduce hyperon potentials. The expansion of the mean field hadronic chiral  non-linear model into quark degrees of freedom (CMF model) shares similarities with the Polyakov loop-extended Nambu-Jona-Lasinio (PNJL) model~\cite{Ratti:2005jh}. The CMF utilizes a scalar field $\Phi$, analogous to the PNJL model, to suppress quark degrees of freedom at low densities and/or temperatures. In our context, $\Phi$ is the scalar field associated with the PNJL-like effective potential that drives the transition from confined to deconfined phases. This transition is phenomenologically captured by the  order parameter  $\Phi\in[0,1]$. The modification of Polyakov loop potential ($U_\Phi$) from its original PNJL model form \cite{Ratti:2005jh, Roessner:2006xn} includes the incorporation of terms dependent on the baryon chemical potential~\cite{Dexheimer:2009hi}. In \cref{eq:U_Phi}), the $a$'s and $T_0$ are parameters fitted to the known constraints of QCD phase diagram at higher temperatures and are discussed in \Cref{sec:results_pol_coupling}. This adaptation enables the exploration of low-temperature and high-density scenarios, such as those encountered in neutron stars. 

% The deconfinement potential in the CMF model reads \cite{Dexheimer:2009hi}
%
% \begin{align}
% U_{\Phi}&=\left(a_0T^4+a_1\mu_B^4+a_2T^2\mu_B^2\right)\Phi^2\nonumber \\ &+a_3T_0^4\ln\left(1-6\Phi^2+8\Phi^3-3\Phi^4\right),
% \label{eq:U_Phi}
% \end{align}
%
The presence of the scalar field $\Phi$  is introduced as an additional contribution to the effective masses of the baryons
\begin{align}
m_i^*&=g_{i\sigma}\sigma+g_{i\zeta}\zeta+g_{i\delta}\delta+\Delta m_i+g_{i\Phi}\Phi^2, 
\label{eq:emh}
\end{align}
and quarks
\begin{align}
m_i^*&=g_{i\sigma}\sigma+g_{i\zeta}\zeta+g_{i\delta}\delta+\Delta m_i+g_{i\Phi}\left(1-\Phi \right).
\label{eq:emq}
\end{align}
In the above equations, $\Phi\sim0$ denotes a system dominated by hadrons, $\Phi\sim1$ represents a quark-dominated state, and intermediate values indicate a coexistence of hadrons and deconfined quarks (relevant only at high temperatures). Moreover, in those equations, the $g's$ are the corresponding coupling constants of fermions with the scalar fields. Note  that the parameter $\Delta m_i$ incorporates effects from additional sources, such as the Higgs field $m^q_0$  (pertaining to quarks), bare mass  $m_0$  (for octet baryons) and   explicit symmetry breaking term $m_3$ (relevant to hyperons) 
\begin{align}
    \Delta m_N &=m_0\,, \nonumber \\
    \Delta m_\Lambda &= m_0  - m_3 \left( \sqrt{2} \sigma_0 + \zeta_0  \right),  \nonumber \\
    \Delta m_\Sigma &= m_0  - m_3 \left( \sqrt{2} \sigma_0 + \zeta_0  \right),  \nonumber \\
    \Delta m_\Xi &= m_0  - m_3 \left( \sqrt{2} \sigma_0 + \zeta_0  \right),  \nonumber \\
      \Delta m_u &=\Delta m_d=m_0^u\,, \quad \Delta m_s =m_0^s\,.
\end{align}

The coupling constants $g's$ between baryons and scalar mesons are fitted in order to obtain correct  masses of the baryons in vacuum. The other parameters ($k$'s and $\epsilon$), related to scalar interactions, are   computed in order to obtain correct vacuum expectation values for  the $\sigma$, $\zeta$, and $\chi$ field equations and to reproduce $\sigma$, $\eta$, and $\eta^{\prime}$ vacuum masses \cite{Dexheimer:2009hi}. In \Cref{tab:scalar_param}, a list of CMF parameters associated with baryons is tabulated, whereas \cref{tab:quark_param} reflects the CMF parameters related to quark sector (the only ones we do not modify in this work).  The quark scalar couplings are fixed as approximately one third of nucleon scalar couplings, whereas vector couplings are set to zero as suggested by Ref.~\cite{Steinheimer:2014kka}. In the next section, we discuss the vector meson interaction Lagrangian ${\cal L}_{\rm vec}$ in detail.

\begin{table}[htp]
\centering
\caption{Parameters related to the scalar interaction for baryons.}
\def\arraystretch{1.8}
\begin{tabular}{ccc}
\hline
\hline
$\sigma_0=-93.3$~MeV&
$\delta_0=0$&
$\zeta_0=-106.56$~MeV\\
$g_{N\sigma}=-9.83$ & 
$g_{N\delta}=-2.34$ & 
$g_{N\zeta}=1.22$\\
$g_{\Lambda\sigma}=-5.52$ & 
$g_{\Lambda\delta}=0$ & $g_{\Lambda\zeta}=-2.3$\\
$g_{\Sigma\sigma}=-4.01$ & $g_{\Sigma\delta}=-6.95$ & $g_{\Sigma\zeta}=-4.44$\\
$g_{\Xi\sigma}=-1.67$ & $g_{\Xi\delta}=-4.61$ & $g_{\Xi\zeta}=-7.75$\\
$k_0=2.37$ & 
$k_1=1.40$ & 
$k_2=-5.55$\\
$k_3=-2.65$ &   
$\chi_0=401.93$~MeV & 
$\epsilon=2/33$\\ 
\hline
\hline
\end{tabular}
\label{tab:scalar_param}
\end{table}

\begin{table}[t!]
\centering
\caption{Parameters related to the scalar  and vector interaction for quarks.}

\begin{tabular}{lcc}
\hline \hline
$g_{q \omega}=0$ & $g_{q \phi}=0$ & $g_{q \rho}=0$ \\
$g_{u \sigma}=-3.00$ & $g_{u \delta}=0$ & $g_{u \zeta}=0$ \\
$g_{d \sigma}=-3.00$ & $g_{d \delta}=0$ & $g_{d \zeta}=0$ \\
$g_{s \sigma}=0$ & $g_{s \delta}=0$ & $g_{s \zeta}=-3.00$ \\
$m^u_0=5$ MeV & $m^d_0=5$ MeV & $m^s_0=150$ MeV\\
\hline \hline
\end{tabular}

\label{tab:quark_param}
\end{table}

\subsection{ Field Redefinition  of the Lagrangian}
\label{sec:Lag_vec_ren}

\subsubsection{Kinetic and mass terms for vector mesons}
\label{sec:kin_mass_vec}

We start with the  simplest scale invariant mass term of  the field redefined   vector interaction Lagrangian denoted by ``tildes''
\begin{equation}
    \mathcal{ \tilde L}_{\rm vec}^{m}=\frac{1}{2} m_V^2  \operatorname{Tr} \tilde{V}_\mu \tilde{V}^\mu,
    \label{eq:inv_vec}
\end{equation}
where $\tilde{V}_\mu$ is the degenerate  field redefined  vector meson matrix given by \Cref{deg_Vmatrix}. Simplifying,
\begin{equation}
    \mathcal{ \tilde  L}_{\rm vec}^{m}=  m^2_V \left(\frac{\tilde \omega^2}{2}+ \frac{\tilde \phi^2}{2} + \frac{3 \tilde \rho^2}{2}+2 \tilde K^*{^2}\right),
\end{equation}
dividing the $\rho$ and $K^*$ terms by  the degeneracy factor 3 and 4, respectively, we obtain
\begin{equation}
    \mathcal{ \tilde  L}_{\rm vec}^{\rm m}=\frac{1}{2} m^2_V \left(\tilde \omega^2+ \tilde \phi^2+ \tilde \rho^2+\tilde K^*{^2}\right).
    \label{eq:massre}
\end{equation}
The equation above suggests that the vector meson nonet is  mass degenerate. To correct that and split the masses, one can add the chiral invariant (CI)\footnote{In \cref{eq:inv_vec_1}, $\mu$ represents Lorentz index whereas $\upmu$ denotes the fit parameter.}~\cite{Papazoglou:1998vr}
\begin{equation}
\mathcal{\tilde L}_{\rm vec}^{\rm CI}=\frac{1}{4} \upmu \operatorname{Tr}\left[\tilde{V}_{\mu \nu} \tilde{V}^{\mu \nu}  \left\langle X  \right\rangle ^2\right],
\label{eq:inv_vec_1}
\end{equation}
where  $\left\langle X  \right\rangle$ is the scalar meson matrix in the mean-field approximation   given by \Cref{Xmatrix_mfa} (for simplicity, we have taken vacuum values of the scalar meson fields),  ${V}_{\mu \nu}$ is the  field redefined  vector meson tensor matrix given by \Cref{deg_V_tenMatrix} and  $\upmu$ is a fit parameter to the vector mesons vacuum mass constraints with mass dimension of negative two~\cite{Papazoglou:1998vr}. In Ref.~\cite{Detlef:Thesis}, an additional invariant term, $\left(\operatorname{Tr} V_{\mu \nu}\right)^2$, was incorporated into the expression presented in Eq.~\eqref{eq:inv_vec_1} to lift the mass degeneracy between the $\rho$ and $\omega$ mesons; however this provides a small correction and  does not offer an explanation for the vector kaon masses. Since the process of renormalizing the vector kaons is a crucial initial step, laying the groundwork for future work beyond mean-field theory, we focus on \Cref{eq:inv_vec_1}.

Expanding it gives
\begin{align}
    \mathcal{\tilde L}_{\rm vec}^{\rm CI}=&\frac{1}{4}\upmu\Bigg(\frac{\sigma_0^2}{2}(\tilde 
 V^{\mu\nu}_\omega)^2 +3 \frac{\sigma_0^2}{2}(\tilde V^{\mu\nu}_\rho)^2+(\tilde 
 V^{\mu\nu}_\phi)^2\zeta_0^2 \nonumber \\
 &+(\tilde 
 V^{\mu\nu}_{K^*})^2(\sigma_0^2+2\zeta_0^2)\Bigg).
\end{align}
Dividing the $\rho$ term by 3 and the $K^*$ term by 4 based on their respective degeneracies, we obtain,
\begin{align}
    \mathcal{\tilde  L}_{\rm vec}^{\rm CI}=&\frac{1}{4}\upmu\Bigg(\frac{\sigma_0^2}{2}( \tilde V^{\mu\nu}_\omega)^2 + \frac{\sigma_0^2}{2}(\tilde  V^{\mu\nu}_\rho)^2+(\tilde V^{\mu\nu}_\phi)^2\zeta_0^2\nonumber \\
    &+\frac{(\tilde V^{\mu\nu}_{K^*})^2}{2}\left(\frac{\sigma_0^2}{2}+\zeta_0^2\right)\Bigg).
\end{align}
The  kinetic energy vector  term  $\mathcal{ L}_{\rm vec}^{\rm kin}=-\frac{1}{4} \operatorname{Tr}\left({V}_{\mu \nu} {V}^{\mu \nu}\right)$   under field redefinition becomes 
\begin{align}
    \mathcal{\tilde L}_{\rm vec}^{\rm kin}&=-\frac{1}{4} \operatorname{Tr}\left(\tilde{V}_{\mu \nu} \tilde{V}^{\mu \nu}\right),\nonumber \\
    &=-\frac{1}{4}\left(\left(\tilde V_\rho^{\mu \nu}\right)^2+\left( \tilde V_{K^*}^{\mu \nu}\right)^2+\left( \tilde V_\omega^{\mu \nu}\right)^2+\left( \tilde V_\phi^{\mu \nu}\right)^2\right).
    \label{eq:ren_kin}
\end{align}

Now, combining the contributions from \cref{eq:inv_vec_1} with \cref{eq:ren_kin} and identifying them to the  old   kinetic energy term

\begin{align}
 \mathcal{ L}_{\rm vec}^{\rm kin}&=  \mathcal{ \tilde L}_{\rm vec}^{\rm kin}+\mathcal{ \tilde L}_{\rm vec}^{\rm CI}, \nonumber \\
&=-\frac{1}{4}\left[1-\upmu \frac{\sigma_0^2}{2}\right]\left(\tilde V_{\tilde \rho}^{\mu \nu}\right)^2 -\frac{1}{4}\left[1-\upmu \frac{\sigma_0^2}{2}\right]\left(\tilde V_{\tilde \omega}^{\mu \nu}\right)^2 \nonumber \\
 &\quad-\frac{1}{4}\left[1-\frac{1}{2} \upmu\left(\frac{\sigma_0^2}{2}+\zeta_0^2\right)\right]  \left(\tilde V_{\tilde K^*}^{\mu \nu}\right)^2 \nonumber \\
 &\quad-\frac{1}{4}\left[1-\upmu \zeta_0^2\right] \left(\tilde V_{\tilde \phi}^{\mu \nu}\right)^2, \nonumber \\
 &= -\frac{1}{4Z_\rho} \left(\tilde{V}_{\tilde{\rho}}^{\mu \nu}\right)^2-\frac{1}{4Z_{\omega} }\left(\tilde{V}_{\tilde{\omega}}^{\mu \nu}\right)^2 \nonumber \\
&\quad-\frac{1}{4Z_{K^*}} \left(\tilde{V}_{\tilde{K^*}}^{\mu \nu}\right)^2
-\frac{1}{4Z_\phi} \left(\tilde{V}_{\tilde{\phi}}^{\mu \nu}\right)^2,
\label{eq:zlag}
\end{align}

we obtain
%
% \begin{align}
%      \tilde{V}_{{\xi}}^{\mu \nu}=\partial^\mu \tilde{\xi}^\nu-\partial^\nu {\tilde \xi}^\mu&=Z_\xi^{1/2}(\partial^\mu {\xi}^\nu-\partial^\nu {\xi}^\mu)=Z_\xi^{1/2} {V}_{{\xi}}^{\mu \nu}, \nonumber \\
%      \mathrm{and}  \quad \quad \quad  \quad \quad  \xi&=Z_\xi^{-1 / 2} \tilde{\xi}\,,  
%     \label{eq:ren_trans}
% \end{align} 

\begin{align}
     \tilde{V}_{{\xi}}^{\mu \nu}=\partial^\mu \tilde{\xi}^\nu-\partial^\nu {\tilde \xi}^\mu&=Z_\xi^{1/2}(\partial^\mu {\xi}^\nu-\partial^\nu {\xi}^\mu)=Z_\xi^{1/2} {V}_{{\xi}}^{\mu \nu},  
    \label{eq:Vren_trans}
\end{align} 
and
\begin{align}
   \tilde{\xi}&=Z_\xi^{1 / 2} \xi\,, 
    \label{eq:Fren_trans}
\end{align} 

where  $\xi=\rho,\omega,K^*,\phi$. Explicitly, the  constants related to field redefinition are given as
%
% \begin{align}
%     & Z_\rho^{-1}=Z_\omega^{-1}=\left(1-\mu \frac{\sigma_0^2}{2}\right), \quad Z_\phi^{-1} =\left(1-\mu \zeta_0^2\right),\nonumber \\
%     &  Z_{K^*}^{-1}=\left(1-\frac{1}{2} \mu\left(\frac{\sigma_0^2}{2}+\zeta_0^2\right)\right).
% \end{align}

\begin{align}
    & Z_\rho=Z_\omega=\frac{1}{\left(1-\upmu \frac{\sigma_0^2}{2}\right)}, \quad Z_\phi =\frac{1}{\left(1-\upmu \zeta_0^2\right)},\nonumber \\
    &  Z_{K^*}=\frac{1}{\left(1-\frac{1}{2} \upmu\left(\frac{\sigma_0^2}{2}+\zeta_0^2\right)\right)}.
\end{align}

The net Lagrangian for the  vector meson fields (with implicit  field redefinition) is evaluated by adding \Cref{eq:massre,eq:zlag} using \Cref{eq:Vren_trans,eq:Fren_trans}
\begin{align}
\mathcal{ L}_{\rm vec}&= \mathcal{ L}_{\rm  vec}^{\rm 
 kin}  +\mathcal{ L}_{\rm  vec}^{\rm  m}, \nonumber \\
&=-\frac{1}{4}\left(\left(V_\rho^{\mu \nu}\right)^2+\left(V_{K^*}^{\mu \nu}\right)^2+\left(V_\omega^{\mu \nu}\right)^2+\left(V_\phi^{\mu \nu}\right)^2\right) \nonumber \\
&\quad +\frac{1}{2} \left(m_\rho^2 \rho^2+m^2_{K^*} K{^*}^2 +m_\omega^2 \omega^2+m_\phi^2 \phi^2\right),
\label{eq:L_vec_kin_m}
\end{align}
where
\begin{align}
 m_{K^*}^2=Z_{K^*} m_V^2,~~ m_{\omega/\rho}^2=Z_{\omega/\rho} m_V^2, ~~m_\phi^2=Z_\phi m_V^2,
\end{align}
denote the respective vector meson masses in the vacuum. The parameters, $m_V=687.33 $ MeV and $\upmu=$ $0.41/\sigma_0^2$ are fitted to obtain  the correct $\omega, \phi$, $\rho$, and $K^*$ masses that are tabulated in \Cref{tab:vec_ren_mass}, together with the   mass without field redefinition.

\begin{table}[t!]
\centering
\caption{Vacuum masses of vector mesons before (old) and  after (new) employing the  field redefinition .}
\begin{tabular}{ccccc}
\hline \hline
 Meson &  $\omega$  & $\rho$& $K^*$&$\phi$    \\ \hline 
Old Mass (MeV) & 687.33  &687.33&687.33&687.33\\ 
New Mass (MeV) &   770.87&770.87&865.89&1007.76\\ 
\hline \hline
\end{tabular}
\label{tab:vec_ren_mass}
\end{table}

\subsubsection{Self-interaction term for vector mesons}
\label{sec:Lag_self_vec_ren}
We start by adding a self-interactive Lagrangian term to \Cref{eq:L_vec_kin_m}
 \begin{align}
\mathcal{ L}_{\rm  vec}&= \mathcal{ L}_{\rm  vec}^{\rm 
 kin}  +\mathcal{ L}_{\rm  vec}^{\rm  m} +\mathcal{ L}_{\rm  vec}^{\rm SI}\,.
\label{eq:L_vec_net}
\end{align}
The different possible  self-interaction (SI) terms of the vector mesons that are chiral invariant~\cite{Dexheimer:2015qha}  can be written as the following coupling schemes: (shown here in  field redefined  version for the first time)
\begin{align}
& \text { $^\mathrm{R}$C1: } \mathcal{\tilde L}_{\rm 
 vec}^{\rm  SI}=2\tilde g_4 \operatorname{Tr}\left(\tilde V^4\right),\\
& \text { $^\mathrm{R}$C2: }\mathcal{\tilde L}_{\rm 
 vec}^{\rm  SI} =\tilde g_4\Bigg[\frac{3}{2}\left[\operatorname{Tr}\left(\tilde V^2\right)\right]^2 -\operatorname{Tr}\left(\tilde V^4\right) \Bigg],\\
& \text { $^\mathrm{R}$C3: }\mathcal{\tilde L}_{\rm 
 vec}^{\rm  SI} =\tilde g_4\left[\operatorname{Tr}\left(\tilde V^2\right)\right]^2,\\
& \text { $^\mathrm{R}$C4: }\mathcal{\tilde L}_{\rm 
 vec}^{\rm  SI}
=\tilde g_4\frac{[\operatorname{Tr}(\tilde V)]^4} {4}\,,
\end{align}
where superscript ``R'' denotes the  field redefined  coupling scheme. The coupling scheme C2,  is a linear combination of C1 and C3 and is constructed in order to eliminate the $\omega \rho$ mixing term. 

Now, after substituting the matrix $V$ (\cref{deg_Vmatrix}) in the above equations and simplifying them,  we obtain the following equations
\begin{widetext}
\begin{itemize}
\item $^\mathrm{R}$C1:
\begin{align*}
\mathcal{\tilde L}_{\rm  vec}^{\rm SI}=\tilde g_4\left(\tilde \omega^4+6 \tilde \omega^2 \tilde \rho^2+ \tilde \rho^4+2 \tilde \phi^4\right),
\end{align*}
\begin{align}
\mathcal{L}_{\rm vec}^{\rm SI}&= \frac{g_4}{Z_\omega^2}\left(Z_\omega^2 \omega^{4}+6 Z_\omega  Z_\rho  \omega^2\rho^2+Z^2_\rho \rho^{4} +2 Z_\phi^2 \phi^{4}\right)= g_4\left( \omega^{4}+6   \frac{Z_\rho}{Z_\omega}\omega^2\rho^2+\left(\frac{Z_\rho}{Z_\omega}\right)^2 \rho^{4} +2 \left(\frac{Z_\phi}{Z_\omega}\right)^2 \phi^{4}\right).
\end{align}
\item $^\mathrm{R}$C2: 
\begin{align*}
\mathcal{\tilde L}_{\rm vec}^{\rm SI}=\tilde g_4\left(\tilde \omega^4 + \tilde \rho^4+ \frac{\tilde \phi^4}{2}+3 \tilde \rho^2 \tilde \phi^2 ++3 \tilde \omega^2 \tilde \phi^2 \right),
\end{align*}
\begin{align}
 \mathcal{ L}_{\rm vec}^{\rm SI}&=\frac{ g_4}{Z^2_\omega}\left(Z^2_\omega  \omega^4 + Z^2_\rho  \rho^4+ Z^2_\phi \frac{ \phi^4}{2} +3  Z_\rho Z_\phi  \rho^2 \phi^2  +3 Z_\omega Z_\phi  \omega^2  \phi^2 \right),\nonumber \\
 &= g_4\left( \omega^4 + \left(\frac{Z_\rho}{{Z_\omega}}\right)^2 \rho^4+ \left(\frac{Z_\phi}{{Z_\omega}}\right)^2  \frac{ \phi^4}{2} +3  \left(\frac{Z_\rho}{{Z_\omega}}\frac{Z_\phi}{{Z_\omega}}\right)  \rho^2 \phi^2  +3 \left(\frac{Z_\phi}{{Z_\omega}}\right) \omega^2  \phi^2 \right).
\end{align}
\item $^\mathrm{R}$C3:
\begin{align*}
\mathcal{ \tilde L}_{\rm vec}^{\rm SI}=g_4\left(\tilde \omega^4+2 \tilde \omega^2 \tilde \rho^2+ \tilde \rho^4+2 \tilde \omega^2 \tilde \phi^2+ \tilde \phi^4 +2 \tilde \rho^2 \tilde \phi^2\right), 
\end{align*}
\begin{align}
 \mathcal{ L}_{\rm vec}^{\rm SI}&=\frac{ g_4}{Z^2_\omega}\left( Z^2_\omega \omega^4+2 Z_\omega Z_\rho  \omega^2  \rho^2+ Z^2_\rho \rho^4+2 Z_\omega Z_\phi  \omega^2  \phi^2 + Z^2_\phi \phi^4 +2 Z_\phi Z_\rho  \rho^2  \phi^2\right),  \nonumber \\
&= g_4\left(  \omega^4+2  \frac{Z_\rho}{Z_\omega}  \omega^2  \rho^2+ \left(\frac{Z_\rho}{Z_\omega}\right)^2 \rho^4+2 \frac{Z_\phi}{Z_\omega} \omega^2  \phi^2 + \left(\frac{Z_\phi}{Z_\omega}\right)^2 \phi^4 +2  \left(\frac{Z_\rho}{{Z_\omega}}\frac{Z_\phi}{{Z_\omega}}\right)  \rho^2  \phi^2 \right).
\end{align}
\item $^\mathrm{R}$C4:
 \begin{equation*}
\mathcal{\tilde L}_{\rm vec}^{\rm SI}=\tilde g_4\left(\tilde \omega^{4}+2\sqrt{2} \tilde \omega^{3} \tilde\phi+3 \tilde\omega^{2} \tilde\phi^{2}+\sqrt{2}\tilde\omega \tilde\phi^{3}+\frac{\tilde \phi^{4}}{4}\right),
\end{equation*}
 \begin{align}
\mathcal{L}_{\rm  vec}^{\rm SI}&= \frac{g_4}{Z_\omega^2}\left(Z_\omega^2 \omega^{4}+2\sqrt{2} Z_\omega^{3/2} \omega^{3} Z_\phi^{1/2}\phi+3 Z_\omega \omega^{2} Z_\phi \phi^{2}+\sqrt{2}Z_\omega^{1/2}\omega Z_\phi^{3/2}\phi^{3}+\frac{Z_\phi^2 \phi^{4}}{4}\right), \nonumber \\
&=g_4\left(\omega^{4}+2\sqrt{2}\bigg(\frac{Z_\phi}{Z_\omega} \bigg)^{1/2} \omega^{3} \phi+3 \bigg(\frac{Z_\phi }{Z_\omega}\bigg)  \omega^{2} \phi^{2}+\sqrt{2} \bigg(\frac{Z_\phi}{Z_\omega}\bigg)^{3/2}\omega \phi^{3}+\frac{1}{4} \bigg(\frac{Z_\phi }{Z_\omega}\bigg)^2\phi^{4}\right),
\end{align}
\end{itemize}
\end{widetext}
which are obtained using the  field redefined  expressions of  the fields in \Cref{eq:Fren_trans} and defining a coupling constant $g_4=Z^2_\omega \tilde{g}_4$. 

The vector coupling constants $g_{N\omega}$, $g_{N\rho}$, and $g_4$ are adjusted to match nuclear  saturation properties, as explained in  \Cref{sec:results_vec_coupling}. Additionally, it is worth noting that the couplings involving interactions between nucleons and $\omega$ mesons, as well as nucleons and $\rho$ mesons, are influenced by the field redefinitions, leading to corresponding  field redefined  coupling constants: $g_{N \omega} \equiv 3 g_V^8 \sqrt{Z_\omega}$ and $g_{N \rho} \equiv g_V^8 \sqrt{Z_\rho}$ \cite{Zschiesche:2003qq}. Furthermore, it is important to highlight that the coupling scheme labeled as $^\mathrm{R}$C4 has a unique characteristic, involving contributions that exhibit linearity with respect to the isoscalar vector field $\phi$, leading to  significant changes in the model's behavior that help to reproduce astrophysical data, such as 2$M_\odot$ neutron stars. 

\section{Results and Discussions}
\label{sec:RnD}

In this section, we present our numerical findings concerning the vector mesons, their masses and the deconfinement potential in the CMF model. These parameters have been adjusted to accurately replicate  experimental data in the realms of low-energy nuclear physics, astrophysics, and first principle theories. In  earlier works, we constrained the CMF model to match  low-energy nuclear physics and astrophysical observations~\cite{Dexheimer:2008ax,Negreiros:2018cho}, as well as lattice QCD results~\cite{Dexheimer:2009hi} available at that time. We have also compared our results with perturbative QCD~\cite{Roark:2018uls}. However, with the emergence of new theoretical methods, techniques, and experiments both on Earth and in space, there has been significant enhancements in the determination of these constraints. In this work, we have leveraged the most up-to-date constraint data extracted from Ref.~\cite{MUSES:2023hyz} and upgraded our model to account for the mass degeneracy of vector mesons. As a result, we have successfully replicated and improved various characteristics within our model associated with different phases or regions of the QCD phase diagram (presented in \Cref{fig:QCDPD}).

In \cref{tab:param_constraints}, we have compiled  the CMF model free parameters, the Lagrangian term they are associated with, and the specific constraints to which they have been calibrated in this work. Note that different parameters affect different constraints (shown in different lines of \cref{tab:param_constraints}). Our table structure only reflects the order in which we chose to fit those parameters.  The  numerical values corresponding to these constraints can be found in their respective sections.

\begin{table*}[t!]
\centering
% \scriptsize
\caption{The free parameters used to fit the constraints in this work.}
\begin{tabular}{ccc}
\hline \hline  Parameter  & Term & Used to constrain \\
\hline
\hline
$g_1^V$, $g_8^V$, $\alpha_V$, $g_4$  &~~~~ $\mathcal{L}_{int}$+$\mathcal{L}_{\rm vec}^{\rm  SI}$~~~~ & $g_{N \phi}=0,$ $g_1^V=\sqrt{6} g_8^V $, $ n_{\mathrm{sat}}$, $B^{\mathrm{sat}} / A$, $E^{\mathrm{sat}}_{\rm sym}$, $ L^{\mathrm{sat}}$, $K$  \\
\hline
$m_V$, $\upmu$ & $\mathcal{L}_{\rm  vec }^{\rm m}$+ $\mathcal{L}_{\rm vec }^{\rm CI}$  & $m_\omega$, $m_\rho$, $m_\phi$ \\
\hline
$m_3$  &$\mathcal{L}_{\rm esb}$ & $U_\Lambda$ \\
\hline  
$a_0$ & & $T^d_c$  \\
$a_1$ & & $n^d_{B,c}$\\
$a_2$ & &$T^{\rm HQ}_c$, $\mu_{B,c}$ \\
$a_3$ & $U_{\Phi}$& $\Phi\in{0,1}$  \\
$T_0 (\mathrm{ pure glue})$ & & $T^d_c$, $\Phi\in{0,1}$ \\
$T_0 (\mathrm{ crossover})$ & & $T^p_c$, $\Phi\in{0,1}$ \\
$g_{q\Phi}$, $g_{B\Phi}$ & & $T^p_c$ \\
\hline \hline
\end{tabular}
\label{tab:param_constraints}
\end{table*}

\subsection{Parameter fitting for the self-interacting vector meson Lagrangian}
\label{sec:results_vec_coupling}

In \cref{tab:value_param_constraints}, we provide the values of the microscopic and macroscopic properties reproduced through the  field redefined  coupling schemes related to the vector sector of the CMF model. Also in \cref{tab:value_param_constraints_hyperon}, we tabulate the values of $m_3$ parameter,  which is fitted to reproduce  hyperon potential for all couplings. Note that the model's scalar sector remains unaltered because it was originally configured to reproduce  vacuum properties. These values have not been significantly updated over time. In contrast, the coupling constants related to the vector sector are configured to reasonably reproduce  constraints coming from nuclear and astrophysical data. Due to the larger amount of freedom in this case we call them ``free''. The vector coupling constants $g_{N\omega}$ and $g_{N\rho}$ represent the interactions of nucleons with the $\omega$ and $\rho$ mean-fields, respectively. We set $g_1^V=\sqrt6g_8^V$ and $\alpha_V=1$ in $g_{N\phi}=\sqrt{\frac{1}{3}} \, g^V_1 - \,\frac{\sqrt2 }{3}g^V_8 \, (4 \, \alpha_V - 1) $, which cancels terms to ensure that nucleons do not couple to the strange meson $\phi$, i.e., $g_{N\phi}=0$. 

Additionally, the parameter $g_4$ denotes the coupling constant for the self-interaction component of the vector  field redefined  Lagrangian. We adjust the values of $g_{N\omega}$ and $g_4$ to reproduce key modern constraints from low-energy nuclear physics for isospin symmetric matter, specifically the nuclear saturation density $n_{\mathrm{sat}}$, binding energy per nucleon $B/A$, and compressibility $K$. These values fall within the ranges of $n_{\mathrm{sat}}=$ 0.14 to 0.17 fm$^{-3}$, $B/A=$ -15.68 to -16.24 MeV, and $K=$ 220 to 315 MeV, respectively. As mentioned earlier, the $^\mathrm{R}$C4 coupling scheme for self-interacting vector mesons stands out from the others due to its linearity with respect to the strange vector meson $\phi$. This distinctive feature requires special treatment compared to other coupling schemes. For example, it introduces a bare mass of $m_0$ = 150 MeV for nucleons to reproduce a lower compressibility, bringing it in better alignment with nuclear physics data.

Conversely, the parameter $g_{N\rho}$ is responsible for the isospin asymmetry within the medium, and it is therefore adjusted to calibrate the model for achieving specific values of the symmetry energy ($E_{\rm sym}$) and the slope parameter ($L$). These values fall within the ranges of $E_{\rm sym}=$ 28.9 to 34.3 MeV and $L=$ 42.16 to 143 MeV, respectively. It is worth noting that the parameters related to compressibility and the slope parameter are not tightly constrained based on current experimental data~\cite{MUSES:2023hyz}. We anticipate more precise constraints from future experiments. In our  current study, we have deliberately chosen the minimum values for $E_{\rm sym}$ and $L$ while maximizing the neutron star mass ($M_{\rm max}\lesssim 2 M_\odot$) and minimizing the radius ($R_{M_{1.4}} \sim$ 13 Km) for hadronic matter, in accordance with observational constraints~\cite{Miller:2021qha,Riley:2021pdl}. 

The parameter $m_3$ plays a crucial role in determining the level of strangeness content in the medium, and its adjustment is carried out to fit the $\Lambda$ hyperon potential ($U_\Lambda$) with value  around $-28$ MeV and reasonable values for the other parameters~\cite{Kumar:2023qcs}. We determine the maximum masses attained by stars generated by each coupling scheme by employing the Tolman-Oppenheimer-Volkoff (TOV) equations \cite{Tolman:1939jz,Oppenheimer:1939ne}.  In order to obtain the correct neutron star radii, it is important to incorporate a distinct EoS that takes into account the proper microphysics for the crust. The crust is necessary below $n_{\rm{sat}}$ because at this point the nuclei becomes more stable than the hadronic degrees of freedom. 
In this study, we opt for the  widely used Baym-Pethick-Sutherland EoS, which encompasses an inner crust, an outer crust, and an atmosphere~\cite{Baym:1971pw}. Note that the calculations for neutron stars include a free Fermi gas of electrons and muons in chemical equilibrium and ensure charge neutrality i.e. $\sum_i n_i Q_i = 0$, where $n_i$ and $Q_i$ are the number density and electric charge of $i^{th}$ particle, respectively.

\begin{table*}[t!]
\centering
\caption{Best fit of free parameters ($m_0$, $g_{N\omega}$,    $g_{N\rho}$ and $g_4$)  for different self-interaction coupling schemes of  field redefined vector mesons   including low-energy nuclear saturation properties ($n_{\mathrm{sat}}$,  $B/A$, $K$,  $E_{\rm sym}$ and $L$) and astrophysics observables ($M_{\rm  max}$, $R_{M_{\rm  max}}$ and  $R_{M_{1.4}}$). The symbol $``*"$ marks the  cases that do not  include hyperons when calculating stellar properties at $T$=0.}
\begin{tabular}{ccccccccccccc}
\hline \hline
 Coupling& $m_0$&$g_{N\omega}$& $g_{N\rho}$& $ g_4$   &   $n_{\mathrm{sat}}$ (fm$^{-3}$) &   $\frac{B}{A}$ (MeV) & $K$ (MeV) &  $E_{\rm sym}$ (MeV)  & $L$ (MeV) & $M_{\rm  max} (M_\odot)$ &$R_{M_{\rm max}}$ (km) &$R_{M_{1.4}}$ (km) \\ \hline
$^\mathrm{R}$C1$^*$&0&13.54&4.77 & 60.66 & 0.151 & -15.76 & 275.70&28.95 & 66.03&1.90 & 11.66& 13.28  \\  
$^\mathrm{R}$C2$^*$&0&13.54& 3.77& 60.66  & 0.151 & -15.76 & 275.70 &28.91 & 89.28&1.98 & 12.14& 13.95 \\ 
$^\mathrm{R}$C3$^*$&0&13.54&4.13& 60.66  & 0.151 & -15.76 & 275.70 &28.92 & 78.97&1.93 & 11.86& 13.60  \\  
$^\mathrm{R}$C4$^*$&150&11.80& 3.98& 43.93   & 0.151 & -15.70 & 303.43 &28.95& 86.42&2.20 & 12.16& 14.07\\ 
$^\mathrm{R}$C4&150&11.80& 3.98& 43.93   & 0.151 & -15.70 & 303.43 &28.95& 86.42&2.16 & 12.07& 13.96\\    
\hline \hline
\end{tabular}
\label{tab:value_param_constraints}
\end{table*}

\begin{table}[t!]
\centering
\caption{The fitted value of parameter $m_3$  for different self-interaction coupling schemes of  field redefined vector mesons}   reproducing the $\Lambda$ hyperon potential ($U_\Lambda$).
\begin{tabular}{ccc}
\hline \hline
 Coupling& $m_3$&$U_{\Lambda} $(MeV) \\ \hline
$^\mathrm{R}$C1&1.256&-27.96 \\  
$^\mathrm{R}$C2&1.256&-27.96 \\ 
$^\mathrm{R}$C3&1.256&-27.96  \\  
$^\mathrm{R}$C4&0.8061&-28.09\\    
\hline \hline
\end{tabular}
\label{tab:value_param_constraints_hyperon} 
\end{table}

\subsection{Parameter fitting for the Polyakov loop-inspired deconfinement potential}
\label{sec:results_pol_coupling}

The CMF model allows us to investigate  strongly interacting systems involving hadrons and/or  quarks. With this approach, we can delve deeply into the processes governing the restoration of chiral symmetry and the occurrence of deconfinement, particularly under conditions of high temperature or density. This versatility allows our formalism to comprehensively explore e.g, hybrid stars, utilizing a single EoS that accommodates various degrees of freedom. In this section, we provide a detailed exploration of the various parameters associated with the deconfinement potential (\cref{eq:U_Phi}) within the CMF model. In \cref{tab:param_constraints}, we listed the free parameters related to $U_\Phi$  which  are meticulously fitted to  reproduce the rigorous theoretical constraints derived from lattice QCD (briefly discussed in the introduction).  The connection between the $U_\Phi$ parameter(s) and the corresponding constraint are mentioned in the following sections. Specific values of these parameters can be found in Table \ref{tab:param_vec_int_pol}. Within our  field redefined  approach, we thoroughly examine the impact of each parameter within their respective following sections, discussing the constraints they are linked to. Our goal is to offer a comprehensive understanding of how these parameters interact with theory and observation, shedding light on the intricate dynamics of the high-energy part of the QCD phase diagram.

\begin{table}[t!]
\centering
\footnotesize
\caption{Summary of free parameters related to deconfinement for different vector couplings, where $g_{B\Phi}=3g_{q\Phi}$. The values of $g_{q\Phi}$ and $T_0$ are given in MeV.}
\begin{tabular}{cccccccc}
\hline \hline
 Coupling& $a_0$ &  $a_1 (10^{-3})$   & $a_2(10^{-3})$& $a_3$  & $g_{q\Phi}$ &$T_0$(glue )& $T_0$(crossover)    \\ \hline
$^\mathrm{R}$C1&-2.50 &-2.05 &- 0.205 &-0.396& 500&292&200\\  
$^\mathrm{R}$C2&-3.00 &-1.95 & -3.90 &-0.396& 490&306&200\\ 
$^\mathrm{R}$C3&-2.75&-2.03&-0.203&-0.396& 500& 299&200\\  
$^\mathrm{R}$C4&-2.45 &-1.81& -36.2&-0.396&470 & 290&200\\
\hline \hline
\end{tabular}
\label{tab:param_vec_int_pol}
\end{table}

\subsubsection{Deconfinement phase transition}

\begin{figure}[t!]
\centering
\includegraphics[scale=0.55]{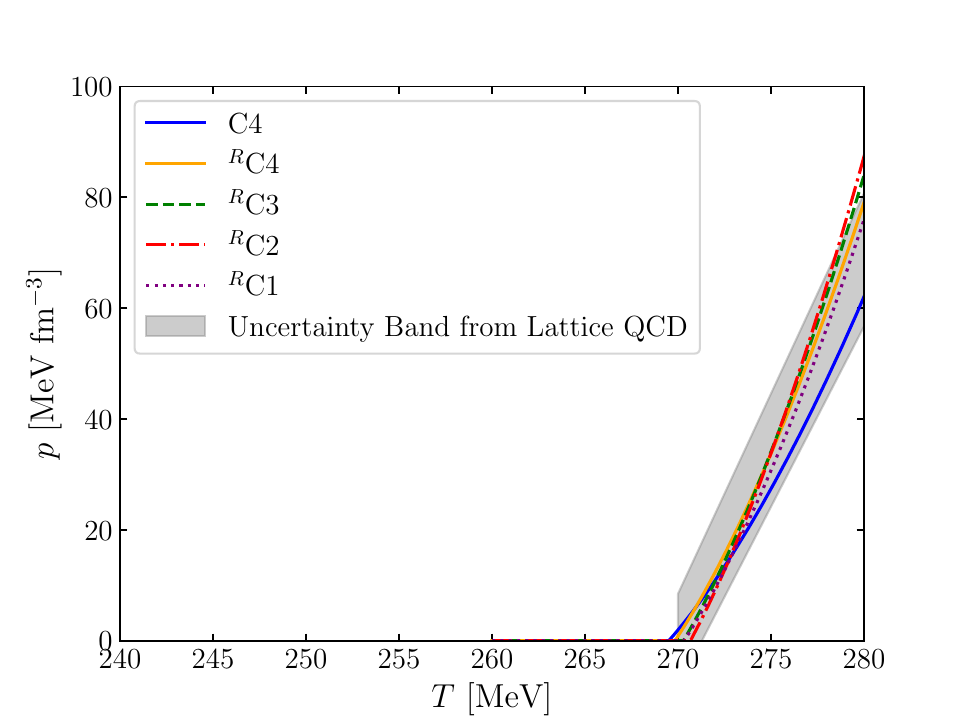}
\caption{Deconfinement phase transition for the  pure glue  case  for different vector couplings with lattice QCD error band taken from Ref.~\cite{Boyd:1996bx}.}
\label{fig:DPT}
\end{figure}

The deconfinement phase transition in QCD is a pivotal shift in the state of matter. It represents the transition from confinement, where quarks and gluons are enclosed within particles like protons and neutrons, to a deconfined state where these fundamental constituents can effectively roam freely. This transition is of paramount importance for understanding the behavior of matter in extreme densities and/or temperatures. In lattice QCD for pure glue  at  $\mu_B=0$, the deconfinement phase transition occurs at $T^d_c \sim 270$ MeV~\cite{Roessner:2006xn}.  In \cref{fig:DPT}, we present the CMF  pressure  for the pure  glue  case compared to lattice QCD calculations. Our CMF results encompass the $^\mathrm{R}$C1-$^\mathrm{R}$C4  field redefined  coupling schemes and one  coupling scheme without field redefinition, as shown in \cref{tab:PCT}. The  coupling scheme C4 (without field redefinition) was the only one for which deconfinement was previously studied and fitted to be  qualitatively similar to the calculations of Refs.~\cite{Ratti:2005jh,Roessner:2006xn}. It is evident that  all couplings lead to a steadily increasing pressure  at temperatures above  the first order phase transition temperature of $T \sim 270$ MeV, indicating that deconfined gluons (in our case, exchange mesons and the field $\Phi$) have a finite  pressure in the deconfined phase. 
 To reproduce the lattice results~\cite{Boyd:1996bx} for $T^d_c$, we perform a parameter fitting for $a_0$ and $T_0$  (refer to Table \ref{tab:param_vec_int_pol} for values) associated with the deconfinement potential, as described in \cref{eq:U_Phi}. All of our parameterizations are within the lattice band for $T\lesssim 280$ MeV.

\subsubsection{Pseudo critical transition temperature}

The chiral phase transition and the deconfinement phase transitions are distinct yet seem to be interconnected phenomena in QCD (at least at $\mu_B=0$). 
The chiral phase transition involves a modification of the QCD vacuum characterized by condensates, crucial for generating hadron masses with chiral symmetry restoration occurring at high temperatures and/or baryonic densities. 
Conversely, the deconfinement phase transition marks the transition from hadronic degrees of freedom to quarks and gluons. 
These transitions are characterized by distinct order parameters (usually $\sigma$ for chiral phase transition and $\Phi$ for deconfinement phase transition). According to lattice QCD findings at $\mu_B=0$, the chiral phase transition from the hadronic phase to the quark phase is not a sharp discontinuity but rather a crossover \cite{Aoki:2006we}. This crossover's central point is denoted as the pseudo-critical or crossover transition temperature $T^p_c$, with a known value of 158 $\pm$ 0.6 MeV as per latest lattice results~\cite{Borsanyi:2020fev}.

\begin{figure}[t!]
\centering
\includegraphics[scale=0.55]{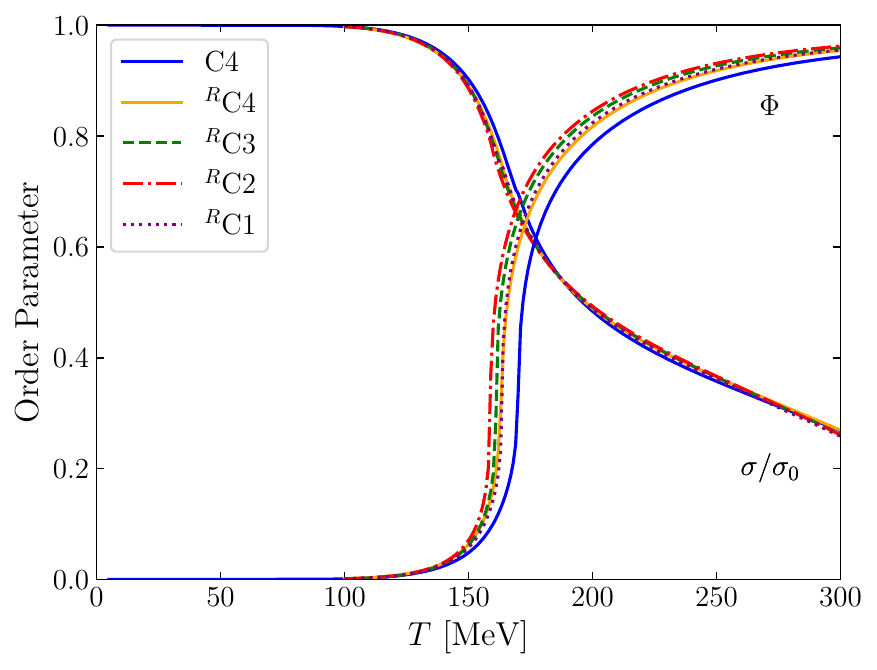}
\caption{Chiral symmetry restoration represented by  the  condensate $\sigma$ and deconfinement represented by $\Phi$ for different couplings at  $\mu_B=\mu_S=\mu_Q=0$.}  
\label{fig:PCT}
\end{figure}

In \cref{fig:PCT}, we present the change in the order parameters $\sigma$ and $\Phi$ with temperature at $\mu_B=\mu_Q=\mu_S=0$.
 In the CMF model, to reproduce the constraints from the theory for $T^p_c$, we perform parameter fitting for $T_0$   and $g_{q\Phi}$,  whose values are  provided in  \cref{tab:param_vec_int_pol}. 
The figure illustrates that the chiral condensate ($\sigma$) is equal to its vacuum value ($\sigma_0$) in the low-temperature regime. However, as  $T$ increases, $\sigma$ decreases, indicating  the transition from the chirally broken phase into the chirally restored phase. Additionally, the maximum change in the chiral condensate (peak of chiral susceptibility) occurs around $T^p_c=$161 MeV  for all  field redefined  coupling schemes, and these values are tabulated in \cref{tab:PCT}. 
Note that, in our model the maximum change in the deconfinement order parameter $\Phi$ is approximately the same as the maximum change in the chiral condensate $\sigma$. For reference, we also mention the value of $T^p_c$ for the older   C4 coupling, which was initially fixed based on the older constraint of the pseudo-critical transition temperature~\cite{Dexheimer:2009hi}, and therefore presents slightly different results. 

\begin{table}[t!]
\centering
\caption{Pseudo-critical temperature for different vector couplings.}
\begin{tabular}{cc}
\hline \hline
 Coupling& $T^p_c$(MeV)  \\ \hline
$^\mathrm{R}$C1& 162.40   \\ 
$^\mathrm{R}$C2& 158.90\\  
$^\mathrm{R}$C3& 161.65  \\ 
$^\mathrm{R}$C4&162.70  \\ 
C4&170.82  \\ 
\hline \hline
\end{tabular}
\label{tab:PCT}
\end{table}

\subsubsection{Deconfinement critical point}

The transition from hadronic to quark phase is characterized by a crossover at low values of baryon chemical potential, but it is believed that eventually a critical point is reached at $\mu_B$, beyond which a first-order phase transition line exists \cite{Stephanov:1998dy}. 
The existence of a critical point is supported by symmetry arguments together with an indication from experiments, where hints of a critical point have been seen in net-proton fluctuation data from the STAR's Beam Energy Scan \cite{STAR:2020tga}.
This phase transition line would intersect the $T=0$ axis at a point a few times the nuclear saturation density. On the theory side,  recent lattice QCD results have not shown  signs of critical behavior up to \mbox{$\mu_B \approx 300$ MeV}, with a critical temperature estimated to be less than {\mbox{$T^{HQ}_c < 132^{+3}_{-6}$ }MeV~\cite{Borsanyi:2020fev,HotQCD:2019xnw}. A machine learning  approach in \cite{Mroczek:2022oga} based on  the lattice QCD equation of state coupled to a critical point  found on the grounds of causality and stability found that the critical point is heavily skewed towards $\mu_B \gtrsim 400$ MeV. To accommodate these constraints, we adjust our model (as provided in \cref{tab:param_vec_int_pol}) to position the critical point at temperatures lower than $T<135$ MeV and baryon chemical potentials greater than $\mu_B>300$ MeV. Note that in a study that used the holographic gauge/gravity correspondence  to map out the QCD phase diagram \cite{Critelli:2017oub}, the authors of \cite{Hippert:2023bel} were able to constrain the location of the critical point at $T_c\sim 105$ MeV and $\mu_{B,c}\sim 580$ MeV by using a Bayesian analysis constrained to state-of-the-art lattice QCD results.

 In our model, to locate the critical point in the region provided by first principles, we make adjustments to the $a_2$ parameter, which is associated with the mixed term $\mu_B^2 T^2$ in the deconfinement potential equation (\cref{eq:U_Phi}). This parameter modification has a direct impact on how the phase diagram behaves in the region where both $\mu_B$ and temperature $T$ are non-zero.  \Cref{fig:PD} illustrates the first-order deconfinement phase transition lines  alongside the respective critical points for various vector coupling schemes. We have also included the phase transition line associated with the  older  C4 coupling scheme, which was fitted to older constraint data. Detailed values of the critical temperature $T_c$, and critical baryon chemical potential $\mu_{B,c}$ for different vector couplings are provided in \cref{tab:_HQCP}. In the cases of $^\mathrm{R}$C1-$^\mathrm{R}$C3, it is notable that the critical point appears naturally at lower values of $\mu_B$ and higher values of $T$ (in comparison with $^\mathrm{R}$C4) no matter how we fix the parameters.

Note that chiral symmetry restoration in the presence of only hadrons appears as a smooth crossover in the CMF model.  When quarks are added, a discontinuity in the order parameter $\sigma$ appears whenever there is a discontinuity  in the order parameter $\Phi$. Nevertheless, we refer to this as a deconfinement phase transition, as the discontinuity in its order parameter is much larger and at low temperature it switches from having just hadrons to just quarks. The overall change in $\sigma$ and, e.g.,  baryon masses (away from the discontinuity) is much more gradual.

\begin{table}[t!]
\centering
\caption{Deconfinement critical point for different couplings.}
\begin{tabular}{ccc}
\hline \hline
Coupling&$T^{\rm HQ}_c$(MeV)& $\mu_{B,c}$ (MeV)\\ \hline
$^\mathrm{R}$C1& 132.0& 1028.85  \\ 
$^\mathrm{R}$C2&127.9 &1042.38\\  
$^\mathrm{R}$C3&132.8&1014.26 \\
$^\mathrm{R}$C4&113.8&1076.39 \\ 
C4&167.0&354.00 \\ 
\hline \hline
\end{tabular}
\label{tab:_HQCP}
\end{table}

\begin{figure}[t!]
\centering
  \includegraphics[scale=0.55]{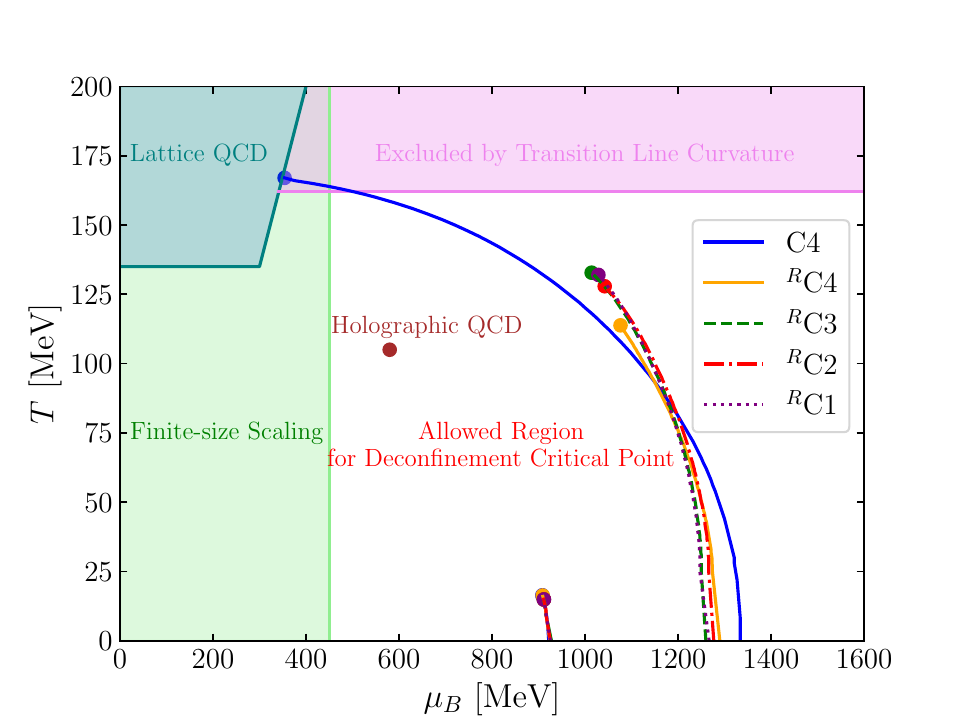}
  \caption{Deconfinement phase transition  as well as liquid-gas coexistence lines and respective critical points for $\mu_Q=0$ and zero net strangeness. The shaded regions show the exclusion of quark hadron critical point by Lattice QCD~\cite{Bazavov:2017dus}, finite-size scaling~\cite{Fraga:2011hi}, and transition line curvature~\cite{Bellwied:2015rza}. It also shows a critical point from holographic QCD~\cite{Critelli:2017oub,Hippert:2023bel}.} 
  \label{fig:PD}
\end{figure}

\subsubsection{Liquid-gas critical point}

In the context of  nuclear physics, the term ``liquid-gas phase transition" is often used to describe a  phase transition akin to what is observed in the behavior of ordinary liquids and gases. In this scenario, the transition  occurs within nuclear matter, which transitions from a phase of nuclei (analogous to a gaseous phase) to bulk nuclear matter (analogous to a more dense liquid phase).  In our model, we do not have nuclei as explicit degrees of freedom, making it a vacuum-to-bulk nuclear matter phase transition. 
Similar to the hadron-quark crossover observed at low baryon chemical potentials, the liquid-gas phase transition also becomes a crossover beyond a threshold temperature. The point that separates the crossover regime from the first-order line is then a  critical point $T^{\rm LG}_c$. 
Beyond this point, it features a distinct discontinuous line known as the liquid-gas phase transition. 

In this study, we have determined for the first time the liquid-gas critical points for various coupling schemes, as documented in \cref{tab:_LGCP}. We do not include hyperons as their influence (if any) would be very small at such ${\mu_B}'s$ and $T's$. We have depicted the liquid-gas phase transition lines for different couplings with critical points in \cref{fig:PD}. This determination is based on the  behavior of the chiral condensate $\sigma$ near $\mu_B \sim 938$ MeV, which corresponds to the mass of nucleons. 6: The liquid-gas critical points were found (without any parameter fitting)  to match experimental observations closely, with values $T^{\rm LG}_{c}$ ranging from 15 to 17 MeV~\cite{Natowitz:2002nw,Karnaukhov:2008be,Elliott:2013pna}. The couplings C1-C3 present slightly different values than C4. This is due to the unique characteristic of C4 involving a linear term in $\phi$ and consequently  different parameterizations including a bare mass term for the baryons.

\begin{table}[t!]
\centering
\caption{Nuclear liquid-gas critical point  for different couplings for $\mu_Q=0$.}
\begin{tabular}{ccc}
\hline \hline
 Coupling& $T^{\rm LG}_c$(MeV)& $\mu^{\rm LG}_{B,c}$ (MeV) \\ \hline
$^\mathrm{R}$C1& 14.91& 911.55  \\ $^\mathrm{R}$C2&14.91 & 911.55\\  
$^\mathrm{R}$C3&14.91 & 911.55 \\ 
$^\mathrm{R}$C4&16.34& 908.94 \\  
C4&16.41&908.32 \\ 
\hline \hline
\end{tabular}
\label{tab:_LGCP}
\end{table}

\subsubsection{Equation of state at T=0}

In this section, we  delve into the $T=0$ axis of the QCD phase diagram, which is approximated by matter in the interior of fully evolved (beyond the proto-neutron star stage)  neutron stars. At  $T=0$, the EoS elucidates
%becomes the focal point, elucidating 
the intricate relationship between various thermodynamic properties of matter within a neutron star and can help to reveal the relevant microscopic degrees of freedom. Leptons (electrons and muons) are included through chemical equilibrium, i.e., $\mu_e=\mu_\mu=-\mu_Q$, where $\mu_Q$ is determined by ensuring electric charge neutrality. $\mu_S$ is set to zero, since strangeness is allowed to increase.

In \cref{fig:EoS_nB}, we  present  pressure versus number density at $T=0$ for four different configurations (one shown also with hyperons). As previously discussed, our model incorporates a deconfinement potential (see Eq.\ \ref{eq:U_Phi}) designed to transition between hadronic and quark contributions. In the figure, for each coupling scheme, we observe an increase in Fermi pressure within the hadronic system as the number density rises, ultimately culminating in a strong first-order  phase transition   (where the horizontal line can be identified with a Maxwell construction, noting that the two extremes in each curve correspond to the same $\mu_B$~\cite{Hempel:2013tfa}). This transition results in a substantial increase in number density as the pressure surges in the quark regime.  Within our newly proposed parametrization, by adjusting the quark couplings to $\Phi$ i.e. $g_{q\Phi}$,  we arrive at a smaller (more realistic) number density jump during the phase transition compared to the old C4 scheme.

 In our quest to gain deeper insights into the threshold of the hadron-to-quark phase transition, characterized by the critical baryonic deconfinement density ($n^d_{B,c}$),  we have adjusted the parameter $a_1$ to obtain a lower value of $n^d_{B,c}$, typically ranging around 3.4 $n_{\mathrm{sat}}$ (compatible with the approximate range of density at which baryons start to overlap). For reference, we have compiled the values of critical densities ($n^d_{B,c}$) obtained within our work in \cref{tab:_PTnB}. Furthermore, the degree of softness or stiffness in the EoS serves as a key determinant of a neutron star's ability to resist gravitational collapse. From the behavior of pressure versus energy density in the quark sector (not shown here), we observe that all of the new coupling schemes exhibit stiffer EoS compared to the old C4 scheme. We also find that in the $^\mathrm{R}$C4 coupling scheme the stiffness (pressure in relation to number/energy density) is almost the same, independently of the presence of hyperons, as they tend to appear in small numbers.  However, the inclusion of hyperons for ($^\mathrm{R}$C1-$^\mathrm{R}$C3) couplings would lead to an extremely soft EoS due to larger number of hyperons. As such scenario is not compatible with recent observations of neutron stars, we chose not to show these results.

\begin{figure}[t!]
\centering
  \includegraphics[scale=0.55]{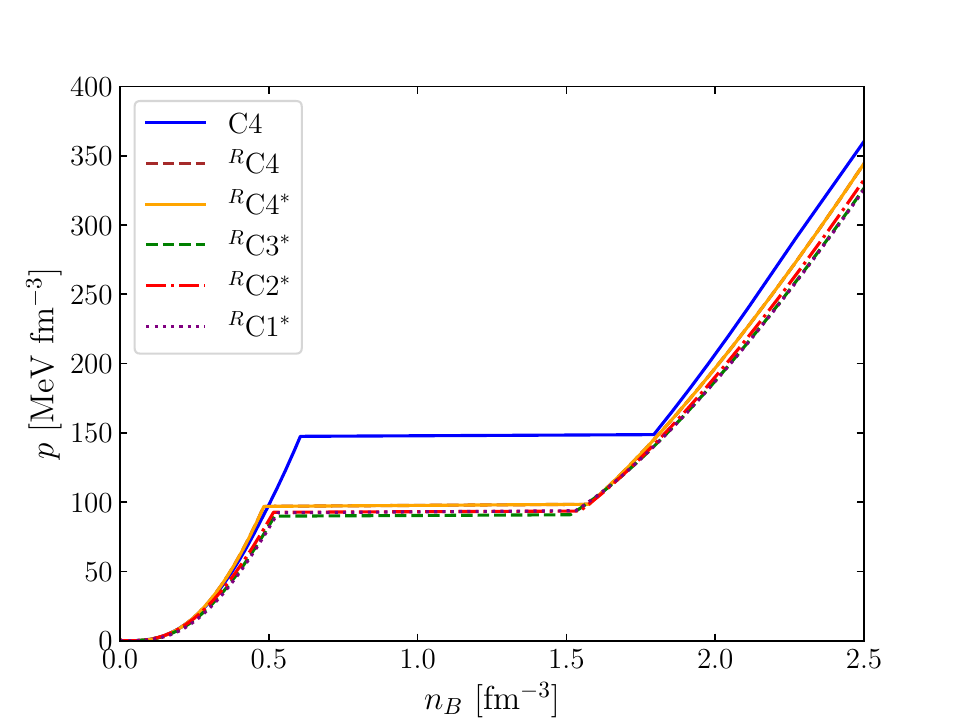}
  \caption{Equation of state  for neutron-star matter at $T=0$ for different vector couplings. The symbol $``*"$ marks the cases that do not include hyperons at $T$=0.}
  \label{fig:EoS_nB}
\end{figure}
\begin{table}[t!]
\centering
\caption{Starting point of deconfinement phase transition for neutron-star matter at $T=0$ using different vector couplings. The symbol $``*"$ marks cases that do not that include hyperons at $T$=0.}
\begin{tabular}{cc}
\hline \hline
Coupling& $n^d_{B,c}(n_{\mathrm{sat}})$ \\ \hline
$^\mathrm{R}$C1$^*$&3.53 \\ 
$^\mathrm{R}$C2$^*$&3.44\\  
$^\mathrm{R}$C3$^*$&3.46 \\ 
$^\mathrm{R}$C4$^*$&3.22  \\
$^\mathrm{R}$C4&3.22  \\ 
C4&4.00 \\ 
\hline \hline
\end{tabular}
\label{tab:_PTnB}
\end{table}

In \cref{fig:MR}, we depict the mass-radius curves for various  field redefined  coupling schemes, both with and without hyperons, within a system governed by hadronic degrees of freedom. To provide context, we also include the mass-radius curve from our widely used work (C4 coupling with hyperons)~\cite{Dexheimer:2008ax,Dexheimer:2009hi,Dexheimer:2011pz,Dexheimer:2015qha,Dexheimer:2018dhb,Dexheimer:2020rlp,Dexheimer:2021sxs,Clevinger:2022xzl,Franzon:2015sya,Marquez:2022fzh,Peterson:2023bmr,Hempel:2013tfa,Roark:2018uls,Aryal:2020ocm,Negreiros:2018cho}. When examining hadronic matter without hyperons, we observe that the  field redefined  $^\mathrm{R}$C4 coupling scheme yields the highest maximum mass ($M_{\rm max}$), which can be compared to other  field redefined coupling schemes and the  old  C4 scheme. The inclusion of hyperons results in a slight reduction in $M_{\rm max}$ for $^\mathrm{R}$C4, but it still remains higher than the other coupling schemes. From the figure, for $^\mathrm{R}$C4 we can conclude that the incorporation of  field redefined vector mesons   leads to a stiffer EoS, resulting in a higher $M_{\rm max}$. On the other hand, for the other  field redefined  coupling schemes ($^\mathrm{R}$C1-$^\mathrm{R}$C3), we also achieve a $M_{max}$ of approximately $2M_{\odot}$.

\begin{figure}[t!]
\centering
  \includegraphics[scale=0.55]{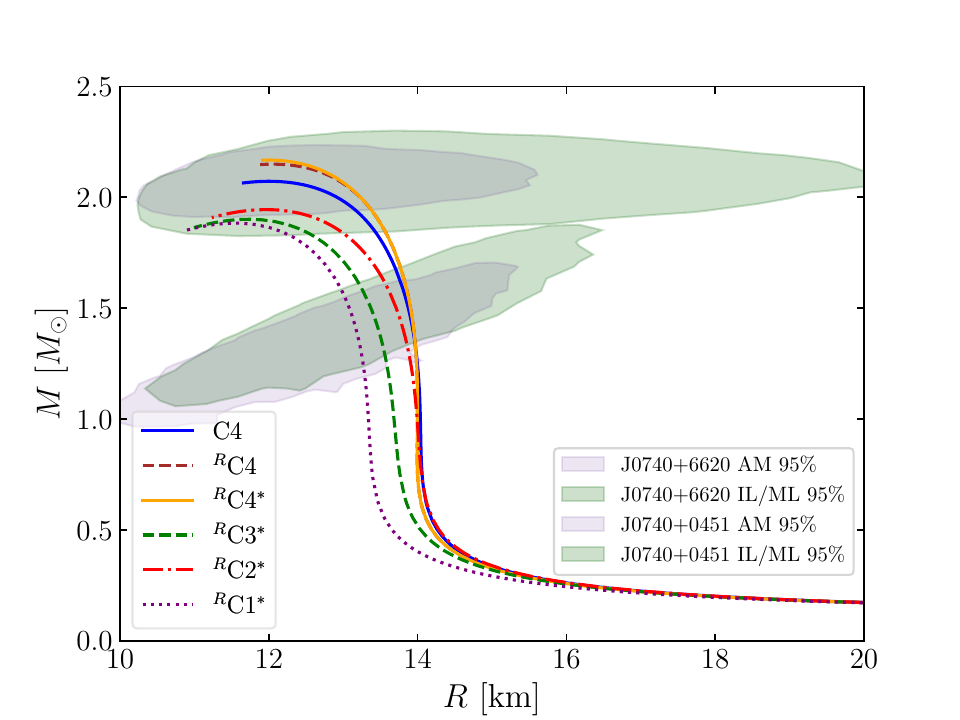}
  \caption{Mass-radius curve for neutron star matter with different  field redefined  vector coupling including a BPS crust~\cite{Baym:1971pw}. The shaded regions in color  green denote the NICER data for PSR J0030+6620 and J0740+0451, employing the Illinois-Maryland analysis~\cite{Miller:2019cac}. Meanwhile, the violet region illustrates the NICER data for the same pulsars analyzed through the Amsterdam analysis~\cite{Riley:2019yda}. The symbol $``*"$ marks the cases that do not include hyperons at $T$=0.}
  \label{fig:MR}
\end{figure}

Concerning radius and by extension tidal deformability, better agreement with the results from NICER~\cite{Miller:2019cac,Riley:2019yda}, LIGO and VIRGO~\cite{LIGOScientific:2018cki} can be achieved by modifying the vector-isovector interactions ($\omega \rho$). This has been explored, e.g., in Refs.~\cite{Horowitz:2002mb,Schramm:2002xa,Bednarek:2013jxa,Steiner:2012rk,Logoteta:2013ipa,Dutra:2014qga,Bizarro:2015wxa,Zhao:2015ncr,Pais:2016xiu,Tolos:2016hhl,Hornick:2018kfi} and in the CMF model~\cite{Dexheimer:2018dhb,Dexheimer:2008ax,Dexheimer:2015qha,Kumar:2023qcs}. In the present work, we do not focus on vector-isovector interactions because they do not modify the finite temperature part of the QCD phase diagram. Also, we did not vary the crust in the present work but a different crust will influence agreement with  LIGO/NICER constraints. 

While we now have a complete equation of state that includes the deconfinement phase transition, a thorough study of its macroscopic properties on neutron stars will wait for a later work. The primary reason is that
including the EoS as it is in the TOV equations implies that the surface tension of quark matter is infinite, which would generate an impenetrable ``wall'' between the hadronic and quark phases. Under the influence of gravity, points with similar pressure would be side by side and there would be no region with the baryon densities corresponding to the jump in \cref{fig:EoS_nB}. In this case, the mass-radius diagram would show a kink. To assess the stability of the star in the decreasing mass branch, we would have to consider the speed of hadron$\leftrightarrow$quark conversion \cite{Lugones:2021zsg}, which could in turn make hybrid stars unstable. Second, if the surface tension of quark matter is below a certain threshold, a mixture of phases appears, which enhances stellar stability. In this case, another dimension ($\mu_Q$) appears, which allows the baryon density to increase smoothly while connecting the hadronic and quark phases \cite{Lugones:2021zsg,Lugones:2021bkm,Chandrasekhar:1964zz,Harrison1965}. Mixed phases have been extensively studied within the CMF model \cite{Hempel:2013tfa,Roark:2018uls} and more will be reported soon.

Alternatively, by altering the deconfined potential (for example, from $a_1 \mu_B^4$ to $a_1' \mu_B^2$) to make it less responsive to the baryon chemical potential, we can achieve a less pronounced first-order phase transitions, resulting in smaller changes in baryon density across the deconfinement phase transition even for infinite surface tension~\cite{Alford:2006bx,Voskresensky:2002hu,Maslov:2018ghi,Wu:2017xaz,Wu:2018zoe,Xia:2020brt}. This facilitates producing stable hybrid stars without mixed phases \cite{Clevinger:2022xzl,Dexheimer:2020rlp,Kumar:2023qcs}. 
Note that quark superconductivity can also influence the mass-radius of neutron stars if the gap size is large enough \cite{Blaschke:2022egm, Ruester:2003zh}, with recent constraints setting a gap limit around a few hundreds of MeV \cite{Kurkela:2024xfh}.

\section{Conclusions}
\label{sec:conclusion}

In this work we take the CMF model, which can describe most key features across the QCD phase diagram, and break degeneracies in the mass of the vector mesons for the first time.  We  explore different self-interaction vector couplings (C1-C4) and for the first time, we study some of them (C1-C3) including deconfinement to quark matter and finite temperature effects. These crucial steps give us a better understanding of the role of vector mesons play in the equation of state and pave the way for future studies of in-medium masses of thermal meson within the CMF model.  Because of the complexity of the CMF model and its inherent interconnectedness across the entire phase diagram, the changes we make also required a full revision of the model.  Furthermore, over the past decade significant advances have been made across the QCD phase diagram.  We incorporate these new constraints in the latest parameterization of the CMF model for the first time in this work as well.

As the vector mesons play a crucial role in mediating the repulsive forces between baryons and quarks, their  field redefinition strongly affects the properties of hadronic and quark matter. Therefore, the entire model needs to be reparameterized. By incorporating appropriate chiral invariants into the vector interactive Lagrangian, we  successfully eliminate the mass degeneracy among the vector mesons by refitting the parameters related to the mass term of the vector meson Lagrangian, aligning them more closely with empirical data.  We  also discuss the fitting of parameters for the baryon/quark-meson interaction and self-interacting vector mesons. These adjustments aim to match key modern experimental constraints, such as the saturation density, the binding energy per nucleon, the compressibility, the symmetry energy,  the slope parameter, and the Lambda hyperon potential in addition to constraints for the liquid-gas critical point and constraints from astrophysics.  In particular, we find that the redefinition of vector fields plays a significant role in reproducing neutron stars with higher masses, when compared to the previous  C4 coupling scheme.

Furthermore, we  explore the parameters associated with the Polyakov loop-inspired deconfinement potential. This includes reproducing lattice QCD  constraints, such as the location of the deconfinement phase transition, the pseudo-critical transition temperature, and constraints that exclude the location of the hadron-quark critical point in certain regions of the phase diagram. The uniqueness of our new parameter fit is grounded in our careful selection of distinctive constants by spanning the search over a whole phase diagram, encompassing novel constraints not previously considered in the field. Rigorous validation, including extensive consistency checks, demonstrates the robustness of our results.

Looking forward, this work opens up multiple new avenues to explore.  For starters, we can study the effect of the new  field redefinition  schemes on the in-medium masses of thermal mesons, which haven't yet been considered. At the moment, we can easily add a gas of free thermal mesons to our calculations, but they would be present in both hadronic and quark phases. Once in-medium masses guarantee that the thermal mesons are suppressed in the quark phase, then, we can use a wider set of lattice QCD results to fit or test our formalism, such as partial pressures \cite{Alba:2017mqu}. On the experimental side, the EoS derived from the CMF model can be used to connect the physics of neutron stars with that of heavy-ion collisions when exploring different isospin and strangeness. 
The EoS at $T=0$ is valuable for gaining insights into both the micro- and macroscopic properties of neutron stars, providing a framework to study, e.g., different net strangeness and quark content in neutron stars, as well as input for simulations of neutron-star cooling and, in the case of finite $T$, input for simulations of neutron star mergers and supernovae. It would be interesting to study these new parametrizations of the CMF EoS in different astrophysical scenarios and also in simulations of heavy-ion collisions. Finally, the knowledge gained on the effect of the parameters across the entire CMF model in terms of how they connect to key features of the QCD phase diagram will play an important role in future work (such as a Bayesian analysis) that uses statistical methods to constrain model parameters.

\section*{Acknowledgements}
We acknowledge support from the National Science Foundation under grants PHY1748621, MUSES OAC-2103680, and NP3M PHY-2116686. We also acknowledge support from the Illinois Campus Cluster, a computing resource that is operated by the Illinois Campus Cluster Program (ICCP) in conjunction with the National Center for Supercomputing Applications (NCSA), which is supported by funds from the University of Illinois at Urbana-Champaign. The authors would like to thank Claudia Ratti for providing comments and assistance in finding the Lattice QCD results.

\appendix
\section{Particle Multiplets}

\label{sec:multiplets}

\begin{itemize}
\item Baryon matrix
\begin{equation}
B=\begin{pmatrix}
\frac{\Sigma^{0}}{\sqrt{2}}+\frac{\Lambda_{0}}{\sqrt{6}} & \Sigma^{+} & p \\
\Sigma^{-} & \frac{-\Sigma^{0}}{\sqrt{2}}+\frac{\Lambda^{0}}{\sqrt{6}} & n \\
\Xi^{-} & \Xi^{0} & -2\frac{\Lambda^{0}}{\sqrt{6}}
\end{pmatrix}
\label{Bmatrix}
\end{equation}

\item Scalar matrix
\begin{equation}
X=\begin{pmatrix}
\frac{\delta^{0}+\sigma}{\sqrt{2}} & \delta^{+} & \kappa^{+} \\
\delta^{-} & \frac{-\delta^{0}+\sigma}{\sqrt{2}} & \kappa^{0} \\
\kappa^{-} & \bar{\kappa}^{0} & \zeta
\end{pmatrix}
\label{Xmatrix}
\end{equation}

\item Scalar matrix in the mean-field approximation
\begin{equation}
\left\langle X  \right\rangle=\begin{pmatrix}
\frac{\delta^{0}+\sigma}{\sqrt{2}} &0& 0\\
 0 & \frac{-\delta^{0}+\sigma}{\sqrt{2}} & 0 \\
0 & 0 & \zeta
\end{pmatrix}
\label{Xmatrix_mfa}
\end{equation}

\item Vector meson matrix
\begin{equation}
V_\mu=\begin{pmatrix}
 \frac{\rho^0_\mu+\omega_\mu}{\sqrt{2}} & \rho^+_{\mu} & K^{*+}_\mu \\
 \rho^-_{\mu} & \frac{-\rho^0_{\mu}+\omega_\mu}{\sqrt{2}} & K^{*0}_\mu \\
 K^{*-}_\mu & \bar{K}^{*0}_\mu & \phi_\mu
\end{pmatrix}
\label{Vmatrix}
\end{equation}

\item Degenerate vector meson matrix
\begin{equation}
V_\mu=\begin{pmatrix}
 \frac{\rho_\mu+\omega_\mu}{\sqrt{2}} & \rho_{\mu} & K^{*}_\mu \\
 \rho_{\mu} & \frac{-\rho_{\mu}+\omega_\mu}{\sqrt{2}} & K^{*}_\mu \\
 K^{*}_\mu & \bar{K}^{*}_\mu & \phi_\mu
\end{pmatrix}
\label{deg_Vmatrix}
\end{equation}

\item Degenerate vector meson tensor matrix
\begin{equation}
V^{\mu\nu} = \begin{pmatrix}
\frac{V^{\mu\nu}_{\rho} + V^{\mu\nu}_\omega}{\sqrt{2}} & V^{\mu\nu}_{\rho} & V^{\mu\nu}_{K^{*}}\\
V^{\mu\nu}_{\rho} & \frac{-V^{\mu\nu}_{\rho} + V^{\mu\nu}_\omega}{\sqrt{2}} & V^{\mu\nu}_{K^{*}}\\
V^{\mu\nu}_{K^{*}} & V^{\mu\nu}_{\bar{K}^{*}} & V^{\mu\nu}_\phi
\end{pmatrix}
\label{deg_V_tenMatrix}
\end{equation}

\end{itemize}

\hspace{0.5cm}

\bibliography{inspire,Not_inspire}
\end{document}